\documentclass[twocolumn, amsmath, amssymb, longbibliography,superscriptaddress ]{revtex4-1} 
\pdfoutput=1

\usepackage{graphicx}
\usepackage{dcolumn}
\usepackage{bm}
\usepackage{bbm}
\usepackage{color}
\usepackage{url}
\usepackage{multirow} 
\usepackage[normalem]{ulem}
\usepackage{xcolor}

\usepackage{epsfig}
\usepackage{color}
\usepackage{amsmath}
\usepackage{amssymb}
\usepackage{longtable}
\usepackage{dcolumn}
\usepackage{bm}
\usepackage{ulem}
\usepackage{float}
\usepackage{mathtools}
\usepackage{xparse}
\NewDocumentCommand{\ceil}{s O{} m}{%
  \IfBooleanTF{#1} 
    {\left\lceil#3\right\rceil} 
    {#2\lceil#3#2\rceil} 
}

\newcommand{\be}{\begin{equation}}
\newcommand{\ee}{\end{equation}}
\newcommand{\bd}{\begin{displaymath}}
\newcommand{\ed}{\end{displaymath}}
\newcommand{\BE}{\begin{eqnarray}}
\newcommand{\EE}{\end{eqnarray}}

\newcommand{\bu}{\ensuremath{\mathbf{u}}}
\newcommand{\bv}{\ensuremath{\mathbf{v}}}

\newcommand{\mcZ}{\mathcal{Z}}

\definecolor{darkgreen}{rgb}{0.0, 0.5, 0.0}

\begin{document}

\preprint{}

\title{Opportunities and challenges for quantum-assisted machine learning in near-term quantum computers}

\author{Alejandro Perdomo-Ortiz}
\email{Correspondence: alejandro.perdomoortiz@nasa.gov}
\affiliation{Quantum Artificial Intelligence Lab., NASA Ames Research Center, Moffett Field, CA 94035, USA}
\affiliation{USRA Research Institute for Advanced Computer Science, Mountain View, CA 94043, USA}
\affiliation{Qubitera, LLC., Mountain View, CA 94041, USA}
\affiliation{Department of Computer Science, University College London, WC1E 6BT London, UK}
\affiliation{Cambridge Quantum Computing Limited, CB2 1UB Cambridge, UK}

\author{Marcello Benedetti}
\affiliation{Quantum Artificial Intelligence Lab., NASA Ames Research Center, Moffett Field, CA 94035, USA}
\affiliation{USRA Research Institute for Advanced Computer Science, Mountain View, CA 94043, USA}
\affiliation{Department of Computer Science, University College London, WC1E 6BT London, UK}
\affiliation{Cambridge Quantum Computing Limited, CB2 1UB Cambridge, UK}

\author{John Realpe-G\'omez}
\affiliation{Quantum Artificial Intelligence Lab., NASA Ames Research Center, Moffett Field, CA 94035, USA}
\affiliation{SGT Inc., Greenbelt, MD 20770, USA}
\affiliation{Instituto de Matem\'aticas Aplicadas, Universidad de Cartagena, Bol\'ivar 130001, Colombia}

\author{Rupak Biswas}
\affiliation{Quantum Artificial Intelligence Lab., NASA Ames Research Center, Moffett Field, CA 94035, USA}
\affiliation{Exploration Technology Directorate, NASA Ames Research Center, Moffett Field, CA 94035, USA}

\date{\today}

\begin{abstract}
With quantum computing technologies nearing the era of commercialization and quantum supremacy, machine learning (ML) appears as one of the promising ``killer'' applications. Despite significant effort, there has been a disconnect between most quantum ML proposals, the needs of ML practitioners, and the capabilities of near-term quantum devices to demonstrate quantum enhancement in the near future. In this contribution to the focus collection on ``What would you do with 1000 qubits?", we provide concrete examples of intractable ML tasks that could be enhanced with near-term devices. We argue that to reach this target, the focus should be on areas where ML researchers are struggling, such as generative models in unsupervised and semi-supervised learning, instead of the popular and more tractable supervised learning techniques. We also highlight the case of classical datasets with potential quantum-like statistical correlations where quantum models could be more suitable. We focus on hybrid quantum-classical approaches and illustrate some of the key challenges we foresee for near-term implementations. Finally, we introduce the quantum-assisted Helmholtz machine (QAHM), an attempt to use near-term quantum devices to tackle high-dimensional datasets of continuous variables. Instead of using quantum computers to assist deep learning, as previous approaches do, the QAHM uses deep learning to extract a low-dimensional binary representation of data, suitable for relatively small quantum processors which can assist the training of an unsupervised generative model. Although we illustrate this concept on a quantum annealer, other quantum platforms could benefit as well from this hybrid quantum-classical framework.
\end{abstract}
\maketitle

\section{Introduction}

With quantum computing technologies nearing the era of commercialization and of quantum supremacy~\cite{Mohseni2017}, it is important to think of potential applications that might benefit from these devices. Machine learning (ML) stands out as a powerful statistical framework to attack problems where exact algorithms are hard to develop. Examples of such problems include image and speech recognition~\cite{krizhevsky2012imagenet,hinton2012deep}, autonomous systems~\cite{levinson2011towards}, medical applications~\cite{esteva2017dermatologist}, biology~\cite{Ching142760}, artificial intelligence~\cite{mnih2015human}, and many others. The development of quantum algorithms that can assist or entirely replace the classical ML routine is an ongoing effort that has attracted a lot of interest in the scientific quantum information community~\cite{neven2009nips, bian2010ising, Denil-2011,wiebe2012quantum, Pudenz-QIP-2013, Lloyd-arXiv-2013, Rebentrost-PRL-2014, wang2017quantum, 2015arXiv151203929Z,Lloyd-NatPhys-2014, schuld2016prediction, Wiebe-arXiv-2015, Benedetti-2016, Benedetti2017, Aaronson-2015,Adachi-arXiv-2015,chancellor2016maximum,Amin-arXiv-2016, kieferova2016tomography, kerenidis2016quantum, wittek2017quantum, Potok2017,Schuld-QML-2015, Romero2017, adcock2015advances, biamonte2016quantum, alvarez2016quantum, lamata2017basic, schuld2017quantum, 2017arXiv170708561C, Benedetti2017b,Benedetti2018,Farhi2018}. We restrict the scope of our perspective to this specific angle and refer to it hereafter as quantum-assisted machine learning (QAML). 

Research in this field has been focusing on tasks such as classification~\cite{Rebentrost-PRL-2014}, regression~\cite{wang2017quantum, wiebe2012quantum, schuld2016prediction}, Gaussian models~\cite{2015arXiv151203929Z}, vector quantization~\cite{Lloyd-arXiv-2013}, principal component analysis~\cite{Lloyd-NatPhys-2014} and other strategies that are routinely used by ML practitioners nowadays. We do not think these approaches would be of practical use in near-term quantum computers. The same reasons that make these techniques so popular, e.g., their scalability and algorithmic efficiency in tackling huge datasets, make them less appealing to become top candidates as killer applications in QAML with devices in the range of 100-1000 qubits. In other words, regardless of the claims about polynomial and even exponential algorithmic speedup, reaching interesting industrial scale applications would require millions or even billions of qubits. Such an advantage is then moot when dealing with real-world datasets and with the quantum devices to become available in the next years in the few thousands-of-qubits regime. As we elaborate in this paper, only a game changer such as the new developments in hybrid classical-quantum algorithms might be able to make a dent in speeding up ML tasks. 

In our perspective here, we propose and emphasize three approaches to maximize the possibility of finding killer applications on near-term quantum computers:

\begin{enumerate}
\item Focus on problems that are currently hard and intractable for the ML community, for example, generative models in unsupervised and semi-supervised learning as described in Sec.~\ref{s:opportunities}.

\item Focus on datasets with potentially intrinsic quantum-like correlations, making quantum computers indispensable; these will provide the most compact and efficient model representation, with the potential of a significant quantum advantage even at the level of 50-100 qubit devices. In Sec.~\ref{ss:qdata} we suggest the case of the cognitive sciences, as a research domain potentially yielding such datasets.

\item Focus on hybrid algorithms where a quantum routine is executed in the intractable step of the classical ML algorithmic pipeline, as described in Sec.~\ref{s:challenges} and Sec.~\ref{s:qahm}.
\end{enumerate}

Each one of these tasks has its own challenges and significant work needs to be done towards having experimental implementations on available quantum hardware (see for example Refs.~\cite{Benedetti-2016, Benedetti2017}). Based on our past experience in implementing QAML algorithms on existing quantum hardware, we provide here some insights into the main challenges and opportunities in this steadily growing research field. Along with illustrations from our earlier work and demonstration on quantum annealers, we attempt to provide general insights and challenges applicable to other quantum computational paradigms such as the gate model.
 
In Sec.~\ref{s:opportunities} we present examples of domains in ML we believe offer viable opportunities for near-term quantum computers. In Sec.~\ref{s:challenges} we present and illustrate the challenges ahead of such implementations and, whenever possible, with demonstrations in real hardware. In Sec.~\ref{s:qahm} we introduce a new and flexible approach, the quantum-assisted Helmholtz machine (QAHM)~\cite{Benedetti2017b}, which has the potential to solve many of the challenges towards a near-term implementation of QAML for real-world industrial scale datasets. In Sec.~\ref{s:summary} we summarize our work.

\section{Opportunities in QAML}\label{s:opportunities}

\subsection{Quantum devices for sampling applications}\label{s:sampling} 

The majority of data being collected daily is unlabeled. Examples of unlabeled data are photos and videos uploaded to the Internet, medical imaging, tweets, audio recordings, financial time series, and sensor data in general. Labeling is the process of data augmentation with task-specific informative tags. But this task is often expensive as it requires humans experts. It is therefore important to design models and algorithms capable of extracting information and structures from unlabeled data; this is the focus of unsupervised ML. But why is this important at all? The discovery of patterns is one of the central aspects of science; scientists do not always know {\it a priori} which patterns they should look for and they need unsupervised tools to extract salient spatio-temporal features from unlabeled data. In general, unsupervised techniques can learn useful representations of high-dimensional data, that have desirable properties such as simplicity and sparsity~\cite{bengio2013representation}. When used in conjunction with supervised techniques such as regressors and classifiers, unsupervised tools can substantially reduce the amount of labeled data required to achieve a desired generalization performance~\cite{erhan2010does}. Connections to reinforcement learning and more specific applications are pointed out in Ref.~\cite{goodfellow2016nips} and references therein. 
 
An unsupervised approach that learns the joint probability of all the variables involved in a problem is often called a {\it generative model}. The name comes from the possibility of inferring any marginal and conditional distribution, which in turns provide a way to generate new data resembling the training set. By following the taxonomy introduced in Ref.~\cite{goodfellow2016nips}, we distinguish generative models as either explicit or implicit density estimators. \textit{Explicit density estimators} are a large family of models which include Boltzmann machines, belief networks, and autoencoders. Depending on the characteristics of the model, they can be learned by variational approximations~\cite{kingma2013auto, rezende2014stochastic, mnih2014neural, sonderby2016ladder, rolfe2016discrete}, Monte Carlo sampling~\cite{ackley1985learning, hinton2006fast, bornschein2014reweighted, bornschein2016bidirectional}, or a combination of both~\cite{salakhutdinov2009deep}. \textit{Implicit density estimators} achieve the generative task indirectly, for example, by casting the problem into a classification task~\cite{bengio2014deep} or finding the Nash equilibrium of a game~\cite{goodfellow2014generative}.
 
Interestingly, generative models with many layers of unobserved stochastic variables (also called hidden variables) have the ability to learn multi-modal distributions over high-dimensional data~\cite{bengio2009learning}. Each additional layer provides an increasingly abstract representation of the data and improves the generalization capability of the model~\cite{hinton2006fast}. These models are usually represented as graphs of stochastic nodes where edges may be directed, undirected, or both. Unfortunately, exact inference is intractable in all but the most trivial topologies~\cite{roth1996hardness}. In practice, learning the parameters of the model is even more demanding as it requires to carry out the intractable inference step for each observed data point and for each learning iteration. The computational bottleneck comes from computing expectation values of several quantities under the complex distribution implemented by the model. Classically, this is approached by Markov chain Monte Carlo (MCMC) techniques that, unfortunately, are known to suffer from the slow-mixing problem~\cite{bengio2013better}. It is indeed difficult for the Markov chain to jump from one mode of the distribution to another when these are separated by low-density regions of relevant size.

\begin{figure}
\includegraphics[width=.49\textwidth]{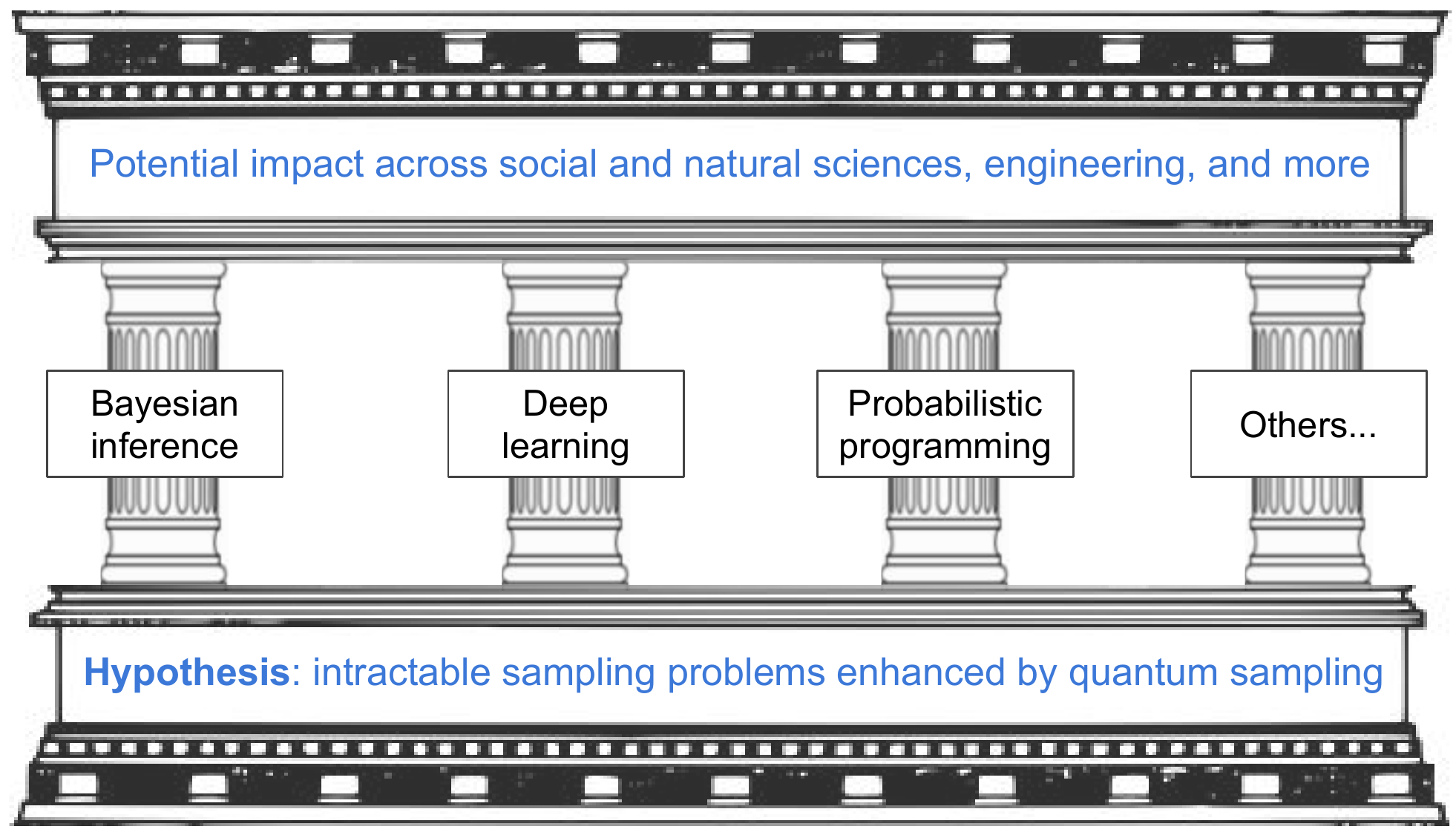}
\caption{\textit{Sampling applications in ML as an opportunity for quantum computers:} Quantum devices have the potential to sample efficiently from complex probability distributions. This task is a computationally intractable step in many ML frameworks, and could have a significant impact on science and engineering.}
\label{f:opportunities}
\end{figure}
 
The capability of quantum computers to efficiently prepare and sample certain probability distributions has attracted the interest of the ML community~\cite{Denil-2011,Dumolin-2014}. In this work, we focus on the use of quantum devices to sample classical~\cite{korenkevych2016benchmarking} or quantum~\cite{Wiebe-arXiv-2015, chowdhury2016quantum} Gibbs distributions as an alternative to MCMC methods. This approach to sampling holds promise for more efficient inference and training in classical and quantum generative models, e.g., classical and quantum Boltzmann machines~\cite{Amin-arXiv-2016}. Once again though, the small number of qubits and the limitations of currently available hardware may impair the sampling process making it useless for real ML applications. In this perspective, we argue that even noisy distributions could be used for generative modeling of real-life datasets. This requires to work in settings where the operations implemented in hardware are only partially known. We call this scenario a {\it gray-box}. We also argue that hybrid classical-quantum architectures are suitable for near-term applications where the classical part is used to bypass some of the limitations of the quantum hardware. We call this approach {\it quantum-assisted}.

As highlighted by ML experts~\cite{LeCun-Nature-2015}, it is expected that unsupervised learning will become far more important than purely supervised learning in the longer term. More specifically, most earlier work in generative models in deep learning relied on the computationally costly step of MCMC, making it hard to scale to large datasets. We believe this represents an opportunity for quantum computers, and has been the motivation of our previous work~\cite{Benedetti-2016, Benedetti2017} and of the new developments presented in Sec.~\ref{s:qahm}.
 
We discussed how inference and learning in graphical models can benefit by more efficient sampling techniques. As illustrated in Figure~\ref{f:opportunities}, sampling is at the core of other leading-edge domains as well, such as probabilistic programming (see Ref.~\cite{lake2015human} for an example of application). If quantum computers can be shown to significantly outperform classical sampling techniques, we expect to see a strong impact across science and engineering.

\subsection{Datasets with native quantum-like correlations}\label{ss:qdata}

Recently, Google demonstrated that quantum computers with as few as 50 qubits can attain quantum supremacy, although in a task with no obvious applications~\cite{boixo2016characterizing}. A highly relevant question then is: Which type of real-life applications could benefit from quantum supremacy with near-term small devices? One of the main motivations underlying the research efforts described here is that quantum computers could speed up ML algorithms. However, they could improve other aspects of ML and artificial intelligence (AI) as well. Recent work shows that quantum mechanics can provide more parsimonious models of stochastic processes than classical models~\cite{Gu2012,palsson2017experimentally,elliott2017occam}, as quantified by an entropic measure of complexity. This suggests that quantum models hold the potential to substantially reduce the amount of other type of computational resources, e.g. memory~\cite{palsson2017experimentally,elliott2017occam}, required to model a given dataset. 

This invites us to pose the following challenge: Identify real-life datasets, unrelated to quantum physics, where quantum models are substantially simpler than classical models, as quantified by standard model comparison techniques such as the Akaike information criterion~\cite{burnham2003model}. In a sense, this is a form of quantum supremacy. While datasets generated in experiments with quantum systems would be an obvious fit, the challenge is to find such data sets elsewhere. Cognitive sciences may offer some potential candidates, as we will describe below.

To avoid potential misunderstandings, let us consider first the example of statistical mechanics, which was developed in the 19th century to describe physical systems with many particles. Although statistical mechanics was long thought to be specific to physics only, we know today that certain aspects of it can be derived from very general information-theoretical principles. For instance, the structure of the Boltzmann distribution can be derived from the maximum entropy principle of information theory~\cite{Jaynes-PhysRev-1957}. The tools of statistical mechanics have become valuable to study phenomena as complex as human behavior~\cite{perc2017statistical,realpe2016balancing}, develop record-performance algorithms~\cite{mezard2002analytic,Mezard-book-2009}, among many other interdisciplinary applications. Indeed, statistical mechanics has played, and continue to play, a relevant role in the development of ML, the Boltzmann machine being an important example. 

In a similar vein, decades of research in quantum foundations and quantum information have allowed the identification of certain features, like interference, entanglement, contextuality, among others, that are encoded naturally by quantum systems~\cite{d2017quantum,realpe2017quantum}. Moreover, there is increasing evidence suggesting that quantum models could be a valuable mathematical tool to study certain puzzling phenomena in the cognitive sciences (e.g., see Refs.~\cite{bruza2015quantum,busemeyer2012quantum,aerts2016quantum} and references therein).

Consider, for instance, the following experiment~\cite{busemeyer2015bayesian}: A participant is asked to play a game where she can either win \$200 or lose \$100 with equal probability, and afterwards she is given the choice of whether or not to play the same gamble again. The experimentalist decides whether or not to tell the participant the result of the first gamble, i.e., whether she won or lost. Results showed that although participants typically preferred to play the second gamble when they knew the outcome of the first gamble, regardless of whether they won or lost, they typically preferred not to play if they did not have such information. More specifically, the participants choose to play the second gamble (G) with probabilities $P(G|W) = 0.69$ or $P(G|L) = 0.59$, respectively, if they knew that they won (W) or lost (L) the first gamble, and with probability $P(G) = 0.39$ if they did not have such information. These results violate the law of total probability $P(G) = P(G|W) P(W) + P(G|L)P(L)$, regardless of the values of the marginal probabilities $P(W)$ and $P(L)$. By interpreting this as an interference phenomenon, the authors have managed to fit the experimental results using a quantum model substantially more parsimonious than alternative classical models, as quantified by standard Bayesian model comparison techniques.

Another general and unexpected prediction is concerned with the effects of the order of questions~\cite{wang2014context}. Asking a yes-no question to a human subject can create a context that affects the answer to a second yes-no question. So the probability of, say, answering yes to both questions depends on which order the questions are asked. For instance, in a 1997 poll in USA, 501 respondents were first asked the question: `Do you generally think Bill Clinton is honest and trustworthy?' Afterwards, they were asked the same question about Al Gore. Other 501 respondents were asked the same two questions, but in reverse order. The number of respondents answering `yes' to both questions significantly increased when Al Gore was judged first. A quantum model for this phenomenon is based on the assumption that the respondent's initial belief regarding the idea of `honest and trustworthy' can be encoded using a quantum state $\rho$. The two possible answers, yes (Y) or no (N), to the question about Clinton or Gore are represented by basis $\{|Y\rangle_C , |N \rangle_C\}$ or $\{|Y\rangle_G , |N \rangle_G\}$, respectively, with respect to which measurements are performed. If the projectors associated to Clinton and Gore do not commute, we have order effects. A general parameter-free equality can be derived from these quantum models for which experimental support has been found in about 70 surveys of about 1000 people each, and two experiments with 100 people~\cite{wang2014context}. Quantum computers may allow for a complementary experimental exploration of these ideas at larger scales, now that experiments with hundreds of people are becoming more common~\cite{wang2014context,gracia2012heterogeneous,gutierrez2014transition}. 

A report~\cite{house2016artificial} from the White House in 2016 reads ``it is unlikely that machines will exhibit broadly-applicable intelligence comparable to or exceeding that of humans in the next 20 years''. Some of the most relevant reasons are technical. Recent work~\cite{realpe2017quantum} suggests why quantum models may be useful in the cognitive sciences and, if so, it may offer new approaches to tackle some of the technical hurdles in AI. The 20-year span predicted above may be in sync with the advent of more powerful and more portable quantum computing technologies. The potential returns may be high.

Although we emphasized the case of cognitive sciences, it would be interesting to explore what other relevant and commercial datasets exhibit quantum-like correlations, and where quantum computers can have an advantage even at the level of 50-100 qubits. In general, the identification of characteristics that are intrinsically quantum, and therefore hard to simulate classically, could be a game changer in the landscape of applications for near-term quantum technologies. Rather than trying to catch up with mature classical technologies in a competition for supremacy, this may offer the opportunity for quantum technologies to create their own unique niche of commercial applications, thereby becoming indispensable.

\section{Challenges in QAML}\label{s:challenges}
 
Near-term implementation of QAML algorithms that can compete with state-of-the-art classical ML algorithms most likely will not come from the quantum version of popular ML algorithms (see e.g., ~\cite{Rebentrost-PRL-2014, wang2017quantum, wiebe2012quantum, schuld2016prediction, 2015arXiv151203929Z, Lloyd-arXiv-2013, Lloyd-NatPhys-2014}). As mentioned in Ref.~\cite{Aaronson-2015}, it would be difficult for these algorithms to preserve the speedup claimed since they inherit the limitations of the Harrow-Hassidim-Lloyd (HHL) algorithm~\cite{harrow2009quantum}. Here, we raise the bar even higher as our attention goes to the implementation of algorithms with potential quantum advantage in near-term quantum devices. 

\begin{figure}
\includegraphics[width=.90\columnwidth]{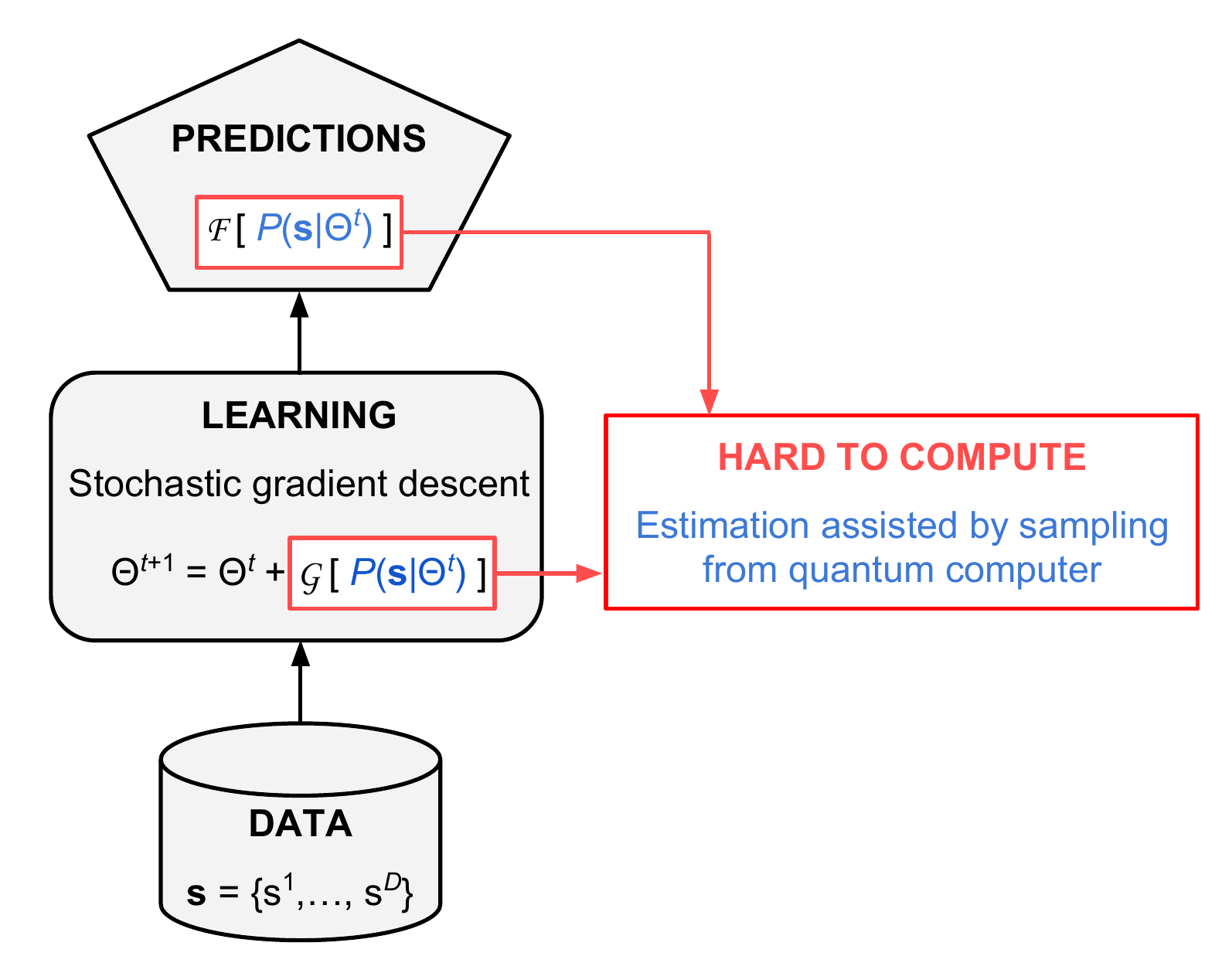}
\caption{\textit{General scheme for hybrid quantum-classical algorithms as one of the most promising research directions to demonstrate quantum enhancement in ML tasks.} A data set drives the fine tuning of model's parameters. In the case of generative models one can use stochastic gradient descent to update the parameters $\Theta$ from time $t$ to $t+1$. The updates often require estimation of an intractable function $\mathcal{G}$ which could be approximated by samples from a probability distribution $P(\mathbf{s}|\Theta^t)$. This computationally hard sampling step could be assisted by a quantum computer. In some cases, making predictions out of the trained model is also an intractable task. The predictions $\mathcal{F}$ could be approximated by samples with the assistance of a quantum computer as well. Examples of such hybrid approaches are illustrated further in Figs.~\ref{f:previous} and~\ref{f:qhelmholtz}.}
\label{f:hybrid}
\end{figure}

As pointed out in Sec.~\ref{s:opportunities}, problem selection is key. For example, consider the recent work in quantum recommendation systems~\cite{kerenidis2016quantum}. The authors developed a custom data structure along with a quantum algorithm for matrix sampling that has polylogarithmic complexity in the matrix dimensions. The result is a quantum recommendation system, and a proposal to circumvents most of the relevant limitations in the HHL algorithm. Because the input size is extremely large (e.g., number of users times number of products), the algorithm promises to significantly speedup the task compared to currently employed ML approaches that require polynomial time. For the very same reasons, however, millions of qubits would be needed to handle datasets where state-of-the-art classical ML starts to struggle. We do not expect such devices to appear in the next decade. Instead, our attention goes to hybrid quantum-classical algorithms where conventional computers are used in the tractable subroutines of the algorithms and quantum computers assist only in the intractable steps. Fig.~\ref{f:hybrid} illustrates an example of this concept for the case of ML tasks. 

There are several challenges which will generally impact any QAML algorithm, such as the limited qubit connectivity, the finite dynamic range of the parameters dictated by the intrinsic energy scale of the interactions in the device, and intrinsic noise in the device leading to decoherence in the qubits and uncertainty in the programmable parameters. We now emphasize some of these practical challenges, with particular attention to execution and implementation of hybrid QAML algorithms in near-term devices.

\begin{figure*}
\includegraphics[width=1.00\textwidth]{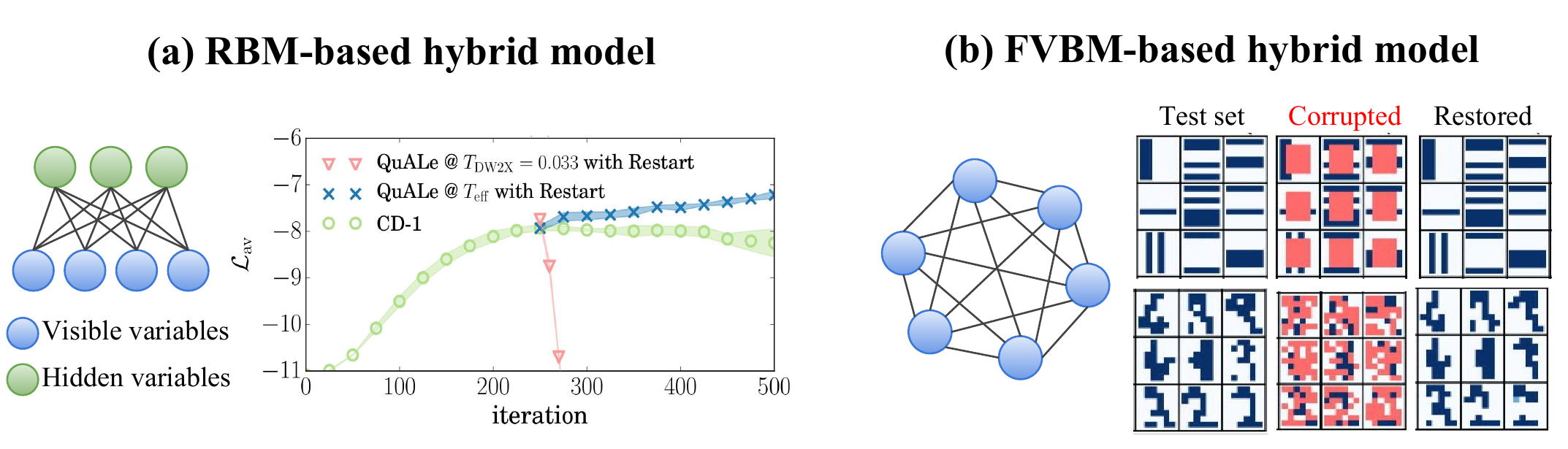}
\caption{\textit{Examples of QAML implementations of different probabilistic graphical models, illustrating some of the challenges in near-term devices.} (a) Restricted Boltzmann machines (RBM) are key components in deep learning approaches such as deep belief networks. Training of these architectures could be significantly improved if one were able to sample from the joint probability distribution of visible and hidden variables, a computationally intractable step usually performed with approximate Markov chain Monte Carlo approaches such as Contrastive Divergence (CD)~\cite{hinton2002training}. Although a quantum annealer can be used to generate such samples, the results need to be combined with those obtained from the classical ML pipeline component within the hybrid approach. This \textit{classical-quantum model compatibility} challenge (Sec.~\ref{ss:cq_match}) is illustrated in the experimental results on the right. $\mathcal{L_{\rm{av}}}$ is the average log-likelihood and represents the performance metric; the higher its value, the better the model is expected to represent the training dataset. As demonstrated in Ref.~\cite{Benedetti-2016}, estimating the instance-dependent effective temperature, $T_{\rm{eff}}$, is key in this QAML approach. In all three lines, the first $250$ iterations are done with the cheapest and widely used version of CD (denoted as CD-1). The line with open circles represents the case where all $500$ iterations in the training are performed with CD-1, and is used as a baseline comparison for our QAML approach. Using our quantum-assisted learning (QuALe) algorithm where $T_{\rm{eff}}$ is estimated at each learning iteration (crosses), we can restart from the point where the classical CD-1 left off and improve the performance of the model with respect to the baseline. Assuming instead that $T_{\rm eff}$ is the physical temperature of the device, $T_{\rm{DW2X}} = 0.033$ (triangles), such restart technique fails. (b) Visible-only generative models are often used either because of their tractability~\cite{goodfellow2016nips} or because of the convexity of the associated optimization problem, as in fully-visible Boltzmann machines (FVBM). In the latter, convexity does not mean tractability; exact learning would still require computation of the partition function which is intractable for nontrivial topologies. Even though there exist fast~\cite{carreira2005contrastive} and consistent~\cite{hyvarinen2006consistency} approximations to the required gradients, we here consider quantum annealing as an alternative tool to sample from nontrivial topologies. In Ref.~\cite{Benedetti2017}, we implemented and trained a hybrid QAML model by introducing a gray-box model (see Sec.~\ref{ss:graybox}) expected to be robust to noise in the programmable parameters and to deviations from the desired Boltzmann distribution, for example, due to non-equilibrium effects in the quantum dynamics. On the right, we show the capabilities of the generative model; two test datasets (leftmost column) are corrupted with different type of noise (red pixels, central column) and then restored on the quantum annealer (rightmost column).}
\label{f:previous}
\end{figure*}

\subsection{Issue of classical and quantum model compatibility}\label{ss:cq_match}

Essential to a hybrid approach is the need to have a flow of information between the classical preprocessing and the quantum experiments. The possibility of sharing information back and forth between the different architectures might pose a significant challenge, arising from the need to match samples from the classical and the quantum model. That is the case, for example, when assisting the training of restricted Boltzmann machines (RBMs) or deep belief networks (DBNs) with quantum devices~\cite{Adachi-arXiv-2015, Benedetti-2016}. There, updates of the parameters are performed by a stochastic gradient descent algorithm that requires two key components. Due to the bipartite structure of RBMs, the first component, also known as the ``positive phase", can be estimated very efficiently with conventional sampling techniques, while the second component, also known as the ``negative phase", is in general intractable and can be assisted with quantum sampling. Since the two terms are subtracted in the same equation and originate from the same model, it is important to have control and to match all the parameters defining both classical and quantum probability distributions. 

A challenge here is related to the temperature of the Gibbs distribution of the model we are sampling from. For simulations in conventional computers, it is irrelevant to explicitly specify the temperature. Since it is a multiplicative factor, we could set it to 1 and ignore it altogether. In the case of an experimental physical device, such as a quantum annealer, we cannot neglect the temperature because (i) it is not under our control; and (ii) it is determined by many factors. These include not only the operational temperature of the device, but also the details of the quantum dynamics. The lack of knowledge of this parameter implies that the communication between classical and quantum components of our hybrid algorithm is broken. In previous work~\cite{Benedetti-2016}, we showed the significant difference in performance that can arise from tackling this challenge. As shown in Fig.~\ref{f:previous}(a), a proper estimation of the temperature also allows us to use restart techniques. For instance, the learning process can initially be carried out on a classical computer because the model parameters may be below the noise level and precision of the quantum device. Then, when desired, the quantum computer can be called to continue the process. 

The challenge can also be addressed at hardware level by designing devices that can prepare a class of quantum Gibbs states at will. However, this strategy might open up other difficulties; some of these are detailed below.

\subsection{Robustness to noise in programmable parameters of the quantum device}\label{ss:graybox}

Preparing quantum Gibbs states with a quantum annealer or with a universal gate model quantum computer is not a straightforward task, given the intrinsic noise in the programmable parameters. In the case of quantum annealers, freezing on quantum distributions or dynamical effects can lead to non-equilibrium distributions away from Gibbs~\cite{Raymond-DWave-2016, korenkevych2016benchmarking}. This is one of the challenges towards scaling the approaches in Refs.~\cite{Adachi-arXiv-2015, Benedetti-2016} where the training of RBMs and DBNs requires reliable samples from a classical Boltzmann distribution. Furthermore, any quantum computing architecture, with intrinsic noise in the programmable parameters, can always lead to deviations from the desired target state. A proposed solution~\cite{Raymond-DWave-2016,korenkevych2016benchmarking} was to use samples obtained from a quantum device to seed classical Gibbs samplers (e.g., MCMC). While this is a promising approach, one of its drawbacks is that it forces the model to have a specific form, even if the quantum device has a genuinely different structure. If the quantum distribution prepared by the device is far from the one assumed, most of the burden would lie on the classical post-processing. It is not clear how much of the method's efficiency is given by the quality of the samples from the quantum devices and how much is achieved by the post-processing steps.

Interestingly, gray-box models such as the fully-visible Boltzmann machine (FVBM) studied in Refs.~\cite{Amin-arXiv-2016,Benedetti2017,korenkevych2016benchmarking} and illustrated in Fig.~\ref{f:previous}~(b) may be a near-term solution to these issues. In a gray-box model, although we assume that samples come from a Gibbs-like distribution, we work directly at the level of the first and second moment statistics, without complete knowledge of the actual parameters that were implemented. Therefore, the emphasis is on the quality of the samples obtained from the quantum device, and their closeness to the data. This has the potential to increase the resilience against perturbations in the programmable parameters. In fact, as long as the estimated gradients for stochastic gradient descent have a positive projection on the direction of the true gradients, the model moves towards the optimum. In cases where this is not so, it might still be possible to design algorithms that mix model-based and model-free information in a suitable way. For instance, a proxy could check whether the estimated gradients actually project on the correct direction; if not, then the system can move in the opposite direction.

Gray-box models with hidden variables could exploit all the available resources from the quantum device, while coping with its intrinsic noise and parameter misspecification. As an example, we used a gray-box approach to implement a quantum-assisted Helmholtz machine where the hidden variables are sampled from a D-Wave device; the framework is discussed in Sec.~\ref{s:qahm} and described in detail in Ref.~\cite{Benedetti2017b}. A caveat of the gray-box approach is that the final model is inevitably tailored for the quantum device used during the training. That is, any time we want to perform ML tasks of interest, such as reconstruction or generation of new images as shown in Fig.~\ref{f:previous}, we will need to use the same quantum device.

\subsection{The curse of limited connectivity}

The basic principle behind this challenge is that physical interactions are local in nature. Although engineering advances can push the degree of qubit connectivity in quantum annealers or gate model quantum computers, required qubit-to-qubit interactions not available in the device will cost an overhead in the computational resources. In the case of quantum annealers, a standard solution is to produce an embedding of the logical problem of interest into the physical layout, therefore increasing the number of required qubits. In the case of a gate model quantum computer, the overhead comes in the number of swaps required to make distant qubits talk to each other~\cite{beals2013efficient}. In any architecture, this compilation requirement needs to be considered in the algorithmic design. 

In the case of quantum annealers, achieving the topological connectivity of the desired model is only half of the challenge. Another significant challenge is the problem of parameter setting associated with the additional interactions present in the embedded model. In other words, how does one set the new parameters such that the embedded model accurately represents the intended model? There are no known optimal solutions, although heuristic strategies have been proposed~\cite{choi2011minor, perdomo2015performance, Pudenz2016}. However, in the type of ML applications we are considering, there is a way out. The main goal in the training phase of a ML algorithm is to find the optimal parameters that minimize a certain performance metric, suggesting that ML itself is a parameter setting procedure. This is precisely the demonstration in Ref.~\cite{Benedetti2017}, where we show how to train models with arbitrary pairwise connectivity. In this case, the difficult task is not the embedding, which can be readily obtained by known heuristics, but rather the training of the whole device, implicitly solving the parameter setting problem. 

To summarize, emphasis has been given to the embedding problem and to the mapping of the logical model of interest into physical hardware. An equally or even more important challenge, is to determine how to set the parameters, including those associated to the embedding, such that the device samples from the desired distribution. This combined problem is what we call \textit{the curse of limited connectivity}. In the case of gate model quantum computer, Ref.~\cite{linke2017experimental} presents an illustrative experimental study, highlighting this challenge. It presents a comparison and analysis of the trade-off between connectivity and quality of computation due to the aforementioned overhead in computational resources. A similar trade-off would need to be taken into consideration in implementations of QAML algorithms in near-term gate model quantum computers.

\subsection{Representation of complex ML datasets into near-term devices}\label{s:complex}

Quantum information does not have to be encoded into binary observables (qubits), it could also be encoded into continuous observables ~\cite{lloyd1999quantum}. There has been work in quantum ML that follows the latter direction~\cite{lau2017quantum, 2017arXiv170700360D}. However, most available quantum computers do work with qubits, nicely resembling the world of classical computation. 
Datasets commonly found in industrial applications have a large number of variables that are not binary. For instance, images may have millions of pixels, where each pixel is a 3-dimensional vector and each entry of the vector is a number specifying the intensity of color. We refer to this kind of datasets as complex ML datasets. A naive binarization of the data will quickly consume the qubits of any device with {100-1000} qubits. Many QAML algorithms~\cite{wiebe2012quantum, Rebentrost-PRL-2014, schuld2016prediction} rely on amplitude encoding instead, a technique where continuous data is stored in the amplitudes of a quantum state. This provides an exponentially efficient representation upon which one could perform linear algebra operations. Unfortunately, it is not clear how one could prepare arbitrary states of this kind on near-term quantum devices. Even reading out all the amplitudes of an output vector might kill or significantly hamper any speedup~\cite{Aaronson-2015}. 

In this perspective, we argue that near-term QAML algorithms should rather aim at encoding continuous variables stochastically into abstract binary representations, a strategy we refer to as {\it semantic binarization}. In this approach, we use quantum states that can be prepared by near-term devices for sampling from unique quantum probability distributions that may capture correlations hard to model with conventional classical ML models. In the context of quantum annealers, such design may allow sampling from non-trivial graph topologies that are usually avoided in favor of restricted ones; for example, bipartite graphs are favored in classical neural networks for convenience.

One way to obtain such an abstract representation is to use hybrid approaches where visible variables $\bv$ are logically implemented by classical hardware, and hidden variables $\bu$ are physically implemented by quantum hardware. However, this idea comes with further challenges. First, the issue of model compatibility described above would apply. Second, sampling hidden variables $\bu$ from the posterior distribution $P(\bu|\bv)$ may be highly problematic because the preparation of arbitrary quantum states is an open challenge. Finally, we might have to sample a binarization for each data point and that would be impractical for near-term quantum computers. For instance, the standard ML dataset of handwritten digits Modified National Institute of Standards and Technology (MNIST) is composed of $60\,000$ training points, hence we would need to program the quantum device at least $60\,000$ times.

As we will see in the next section, deep learning may provide solutions to these challenges. We propose a new hybrid quantum-classical paradigm where the objective is to tackle most of the issues discussed here and, at the same time, to deal with complex ML datasets.

\section{The Quantum-Assisted Helmholtz Machine}~\label{s:qahm}

The quantum-assisted Helmholtz machine (QAHM) is a framework for hybrid quantum-classical ML with the potential of coping with real-world datasets. We already pointed out some of the challenges in developing near-term QAML capable of competing with conventional ML pipelines. Most importantly, the encoding of continuous variables, the limited number of variables, the need to prepare and measure quantum states for each data point. Here we show how some early ideas from the deep learning community can help avoid some of these difficulties. Details about the formalism and preliminary implementation of QAHM can be found in Ref.~\cite{Benedetti2017b}. 

\begin{figure*}
\centering
\raisebox{-0.5\height}{\includegraphics[width=.58\textwidth]{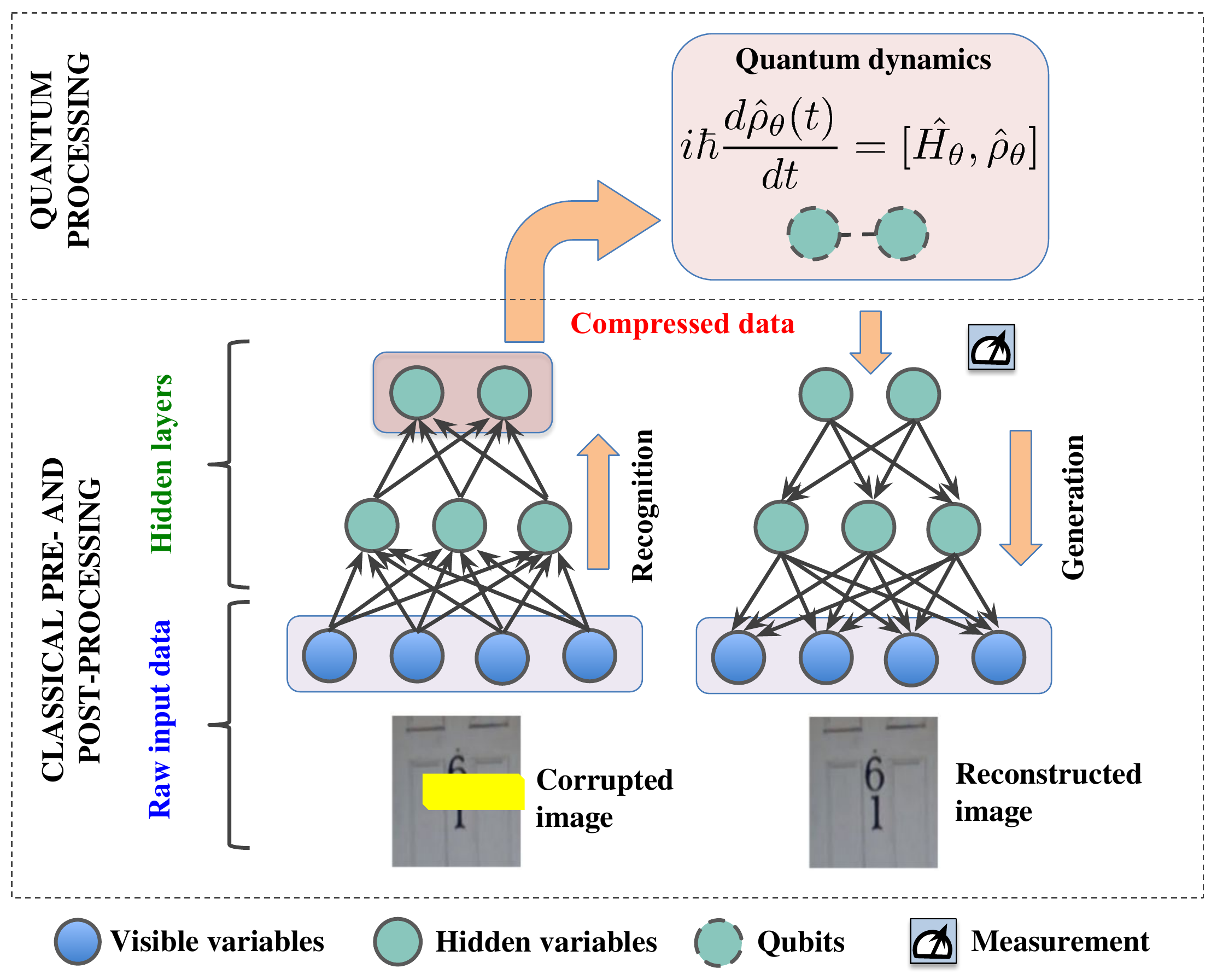}}
\hspace*{.2in}
\raisebox{-0.5\height}{\includegraphics[width=.35\textwidth]{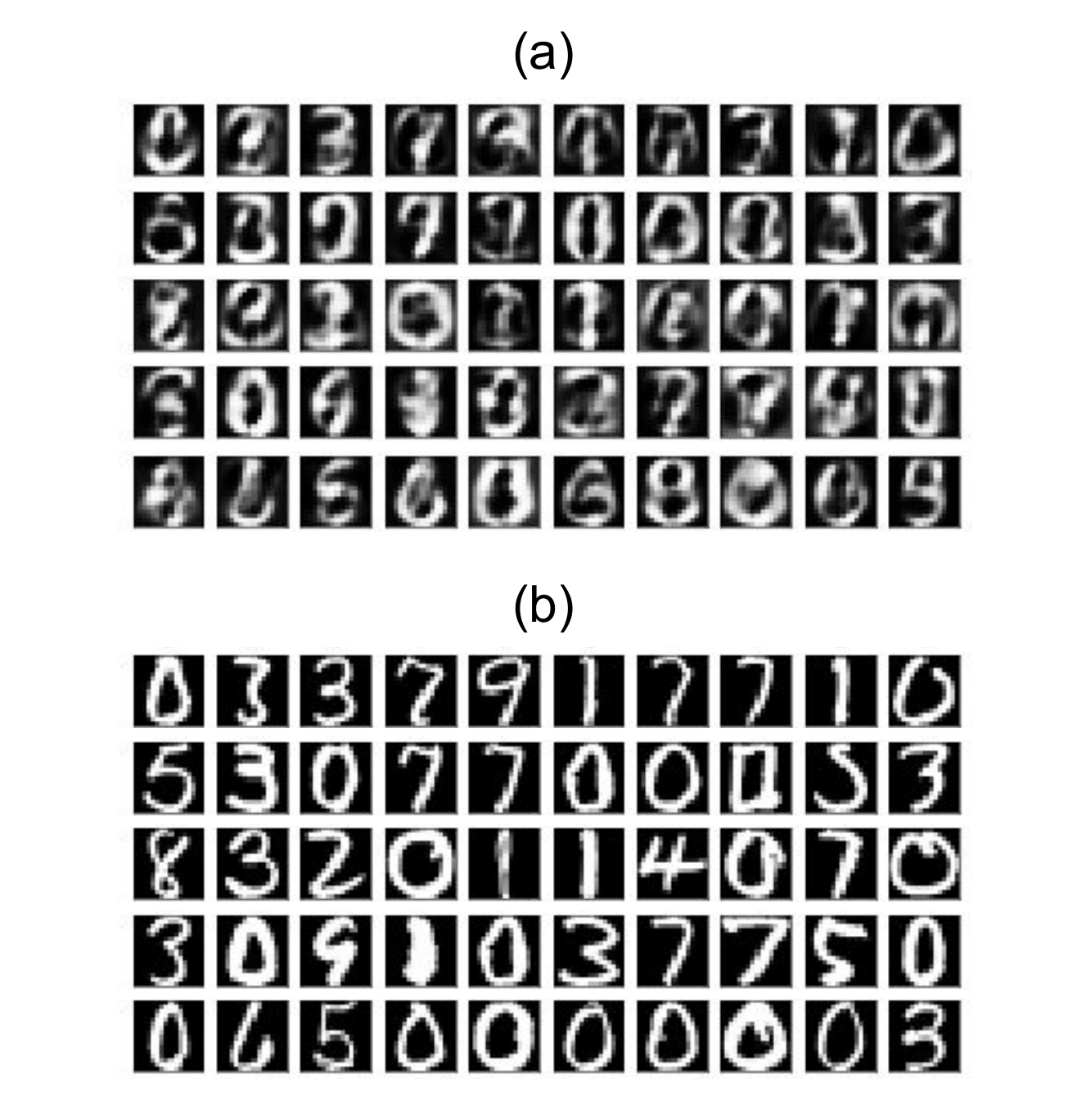}}
\caption{\textit{Generation of handwritten digits with a type of quantum-assisted Helmholtz machine (QAHM).} (Left) The QAHM is the framework we propose to model complex ML datasets in near-term devices. By complex, we refer to datasets where the number of variables is much larger than the number of qubits available in the quantum device, and where data may be continuous rather than discrete. The framework employs a quantum computer to model the deepest hidden layers, containing the most abstract representation of the data. This low-dimensional compact representation is where we believe the quantum device can capture non-trivial correlations and where quantum distributions might have a significant effect. The number of hidden variables in the deepest layers is much smaller than the number of visible variables, making it ideal for near-term implementation on early quantum technologies either on quantum annealers, or universal gate-model quantum computers. $\theta$ indicates the parameters of the quantum computer to be learned, and control the samples obtained from it. Although we illustrate in panel~(a) and ~(b) the realization on a quantum annealer from Ref.~\cite{Benedetti2017b}, extensions to gate model quantum computers are in progress. (a) Artificial data generated by a QAHM implemented on the D-Wave 2000Q, trained on a sub-sampled version of the MNIST dataset with $16 \times 16$ continuous valued pixels and $10$ binary variables indicating the class in $\{0,\dots,9\}$. Both recognition and generator networks have $266$ visible variables and two layers of $120$ and $60$ hidden variables, respectively. The samples are generated from the final model by first sampling the deepest layer with the D-Wave 2000Q, and then transforming those samples through the classical part of the generator network. These experiments use $1644$ qubits of the D-Wave 2000Q quantum annealer. Some of the samples resemble blurry variations of digits written by humans; this problem affects other approaches as well. (b) Samples from the MNIST dataset that are closest in Euclidean distance to those generated by our model. The model does not simply memorize the training set, but rather reproduce its statistics. In future work, the QAHM will be fine-tuned to provide sharper results.}
\label{f:qhelmholtz}
\end{figure*}

Consider a dataset and the task of modeling its empirical distribution with a generative model ${P(\bv) = \sum_{\bu} P(\bv|\bu) P_{QC}(\bu)}$ (see Sec.~\ref{s:sampling} for a brief introduction). Here $\bv$ are the visible variables that represent the data and $\bu$ are unobservable or hidden variables that serve to capture non-trivial correlations. It is common to use binary valued stochastic hidden variables as they can express the presence or absence of features in the data. The set of hidden variables can be partitioned into a sequence of layers that encode increasingly abstract features. In other words, $P(\bv)$ is a deep neural network, called {\it generator network}. 

Here we suggest using a quantum device to model the most abstract representation of the data, that is, the deepest layers of the generator network. The samples obtained from a quantum device are described by the diagonal elements of a parametrized density matrix, $P_{QC}(\bu) = \langle\bu|\hat{\rho}|\bu\rangle$. As an example, the density could be parametrized by a quantum Gibbs distribution $\hat{\rho} = e^{-\beta H}/\mcZ$, where $H$ is the Hamiltonian implemented in quantum hardware and $\mcZ$ is the corresponding partition function. The conditional distribution $P(\bv|\bu)$ is then a classical neural network that transforms samples from the quantum device into samples with the same structure of those in the dataset. Hence, visible variables $\bv$ could be continuous variables, discrete variables, or other objects, effectively tackling the challenge of representing complex data (see Sec.~\ref{s:complex}). Because the quantum device works on a lower dimensional binary representation of the data, this model is also able to handle datasets whose dimensionality is much larger than the number of qubits available in hardware.

Typically, learning algorithms for generative models attempt to maximize the average likelihood of the data. As pointed out in Sec.~\ref{s:sampling}, this is not feasible in models with multiple layers of discrete hidden variables as we would need to sample from the intractable posterior distribution $P(\bu|\bv)$. A Helmholtz machine~\cite{hinton1995wake, dayan1995helmholtz} consists of a generator network along with a {\it recognition network} $Q(\bu|\bv)$ that learns to approximate $P(\bu|\bv)$. This is a key approach behind many variational learning and importance sampling algorithms employed nowadays ~\cite{kingma2013auto, rezende2014stochastic, mnih2014neural, sonderby2016ladder, rolfe2016discrete, hinton2006fast, bornschein2014reweighted, bornschein2016bidirectional, salakhutdinov2009deep}. In principle, the recognition network can also be implemented as a deep neural network whose hidden layers are modeled by a quantum device. However, this design requires quantum state preparation and measurement of the hidden variables for each data point in the training set and for each learning iteration, a daunting process in near-term implementations. To avoid doing this, we will implement the recognition network as a classical deep neural network. Restricting the recognition network to be classical is not a feature of the QAHM framework; instead it is an option to speed up learning of large datasets assisted by serial quantum devices (e.g., quantum annealers). To force the recognition network $Q(\bu|\bv)$ to be close to the true posterior $P(\bu|\bv)$ a notion of distance between them is minimized at each learning iteration. Using the Kullback-Leibler divergence, for instance, leads to the so-called wake-sleep algorithm~\cite{hinton1995wake, dayan1996varieties, bornschein2014reweighted}. 

The general architecture of a type of QAHM is illustrated in Fig.~\ref{f:qhelmholtz}. The recognition network (left) infers hidden variables via a bottom-up pass starting from the raw data. The most abstract representation is obtained either from a classical layer (near-term) or from a quantum device (future implementations). The generator network generates samples of the visible variables via a top-down pass starting from samples obtained from a quantum device. The final model is an implicit density model when a gray-box approach is used to characterize the quantum hardware (see Sec.~\ref{ss:graybox}), but we can turn it into an explicit density model if further processing is employed (e.g., quantum annealing to seed Gibbs samplers). Tasks such as reconstruction, generation, and classification can also be implemented in the QAHM framework.

In Ref.~\cite{Benedetti2017b}, we tested these ideas using a D-Wave 2000Q quantum annealer for the generation of artificial images. For this task we used a sub-sampled $16 \times 16$ pixels version of the standard handwritten digit dataset MNIST. Each gray-scale pixel is characterized by an integer value in $\{0,\dots, 255\}$; we rescale this value in the range $[-1,+1]$ and interpret it as a continuous variable. There are also $10$ binary variables indicating membership to one of the classes. The $266$ visible variables needed to encode this data could, in principle, be embedded directly on the D-Wave 2000Q using a FVBM and a much poorer representation of the data via a naive binarization. Yet we would have to choose a relatively sparse model topology as we cannot embed an all-to-all connectivity in the D-Wave 2000Q for the 266 variables. The sparse connectivity and the absence of hidden variables can severely limit the ability to model the dataset. Our approach tackles both challenges and also enables the handling of larger datasets than would be possible in state-of-the-art quantum annealers. 

We used a classical recognition network and a quantum-assisted generator network, both with $266$ visible variables and two hidden layers of $120$ and $60$ variables, respectively. The deepest layer of $60$ variables was mapped to $1644$ qubits in D-Wave 2000Q using the approach in Ref.~\cite{Benedetti2017}. We run the wake-sleep algorithm for $1000$ iterations and generated samples with the quantum-assisted generator network. The images generated are shown in Fig.~\ref{f:qhelmholtz}(a). Although these preliminary results cannot compete with state-of-the-art ML, the artificial data often resemble digits written by humans. Figure~\ref{f:qhelmholtz}(b) shows the images in the training set that are closest in Euclidean distance to the generated samples. We can see that the artificial images generated by the network are not merely copies of the training set; instead, they present variations and novelty in some cases, reflecting the generalization capability of the model. While the artificial data may also looks blurry, this problem affects other approaches as well. Only the recent development of GANs~\cite{goodfellow2014generative} led to much sharper artificial images. 

\section{Summary}\label{s:summary}

Machine learning (ML) has been presented as one of the application with commercial value for near-term technologies. However, there seems to be a disconnect between the quantum algorithms proposed in much of the literature and the needs of the ML community. While most of the quantum algorithms for ML show that quantum computers have the potential of being very efficient at doing linear algebra (e.g., ~\cite{harrow2009quantum, Rebentrost-PRL-2014, wang2017quantum, wiebe2012quantum, schuld2016prediction, 2015arXiv151203929Z, Lloyd-arXiv-2013, Lloyd-NatPhys-2014}), as discussed in Ref.~\cite{Aaronson-2015}, these proposals do not address the issues related to any near-term implementation. More importantly, to date there are no concrete benchmarks indicating that such work can be close to outperforming their conventional classical ML counterparts. In this perspective we stress this disconnect and provide our views on key aspects to consider towards building a robust readiness roadmap of QAML in near-term devices.

If a demonstration of quantum advantage on industrial applications is a first milestone to be pursued, we emphasize the need to move away from the popular and tractable ML implementations. We should rather look for applications that are highly desirable, but not-so-popular because of their intractability. It is in this domain where we believe quantum computers can have a significant impact in ML. In that sense, quantum speedup in itself is not enough; if the chosen ML applications are tractable to a great accuracy with classical ML methods (as it is the case of Ref.~\cite{kerenidis2016quantum}), then the number of qubits required to tackle industrial-scale applications may be far larger than those available in near-term devices. 

As an example of intractable applications, in Sec.~\ref{s:opportunities} we presented the case of sampling from complex probability distributions with quantum devices. There is potential here for boosting the training of generative models in unsupervised or semi-supervised learning. Another approach we suggested is to explore applications where quantum distributions naturally fit the model describing the data correlations. That seems to be the case for some datasets from the field of cognitive sciences (see e.g., Refs.~\cite{wang2014context,kvam2015interference,busemeyer2015bayesian,realpe2017quantum}; also~\cite{bruza2015quantum,busemeyer2012quantum,aerts2016quantum} and references therein). Other hard ML problems have been mentioned elsewhere~\cite{2017arXiv170708561C}. We think working in any of these currently intractable applications will yield a higher payoff towards demonstrating that quantum models implemented in near-term devices might surpass models trained with classical resources. 

Here we focused on some of the most pressing challenges we foresee in near-term implementations. For instance, the limited qubit connectivity will result in an overhead of qubits in the adiabatic model, and an overhead of gate operations in the gate model. It is also important to take into consideration the model complexity each physical hardware might present, which could have significant consequences on the ML task. For instance, applications to cognitive sciences might require a universal quantum computer capable of preparing and fine-tuning tailored quantum distributions, while applications to generative modeling might only need sampling from a quantum Gibbs distribution. For the latter, there already exist proposals with both quantum annealers and gate model quantum computer architectures. 

Coping with the challenges presented in Sec.~\ref{s:challenges} is certainly an ongoing research activity. One key strategy proposed here towards the near-term demonstration of quantum advantage is the development of hybrid quantum-classical algorithms capable of exploiting the best of both worlds. In this perspective, we also put forward a new framework for such hybrid QAML algorithms, referred here as the \textit{quantum-assisted Helmholtz machine}~\cite{Benedetti2017b}. This new approach aims to solve some of the most pressing issues towards handling industrial-scale datasets with a large number of continuous variables. It is motivated by the idea that a quantum computer should be used only to tackle the more abstract representation of the data, after trimming the information that can be handled classically. Here we use a deep neural network to transform the large continuous data into a new abstract discrete dataset with reduced dimensionality. It is to this abstract representation that we apply the quantum computer. This approach can be used to solve practical ML tasks, including reconstruction, classification and generation of images.

Certainly, more work is needed to address the question of identifying the first killer ML application that can be implemented in near-term quantum computers with the order of a few thousand qubits. Finding intractable problems and developing new hybrid QAML algorithms which tackle the challenges of working with real-world devices, is what we find the ideal scenario. We hope the community makes a leap in this direction now that more powerful and larger quantum annealers and gate model quantum computers are becoming available to the scientific community. 

\section*{Acknowledgements}

The work of A.P-O., J-R-G, and M.B.~ was supported in part by the AFRL Information
Directorate under grant F4HBKC4162G001, the Office of the Director of
National Intelligence (ODNI), the Intelligence Advanced Research
Projects Activity (IARPA), via IAA 145483, and the U.S. Army TARDEC under the ``Quantum-assisted Machine Learning for Mobility Studies" project.
The views and conclusions contained herein are those of the authors and
should not be interpreted as necessarily representing the official
policies or endorsements, either expressed or implied, of ODNI, IARPA, AFRL, U.S. Army TARDEC or the U.S. Government. The U.S. Government is authorized to
reproduce and distribute reprints for Governmental purpose
notwithstanding any copyright annotation thereon. M.B. was partially supported by the UK Engineering and Physical Sciences Research Council (EPSRC) and by Cambridge Quantum Computing Limited (CQCL). The authors would like to thank Leonard Wossnig, Jonathan Romero, and Max Wilson for useful feedback on an early version of this manuscript.



\begin{thebibliography}{100}%
\makeatletter
\providecommand \@ifxundefined [1]{%
 \@ifx{#1\undefined}
}%
\providecommand \@ifnum [1]{%
 \ifnum #1\expandafter \@firstoftwo
 \else \expandafter \@secondoftwo
 \fi
}%
\providecommand \@ifx [1]{%
 \ifx #1\expandafter \@firstoftwo
 \else \expandafter \@secondoftwo
 \fi
}%
\providecommand \natexlab [1]{#1}%
\providecommand \enquote  [1]{``#1''}%
\providecommand \bibnamefont  [1]{#1}%
\providecommand \bibfnamefont [1]{#1}%
\providecommand \citenamefont [1]{#1}%
\providecommand \href@noop [0]{\@secondoftwo}%
\providecommand \href [0]{\begingroup \@sanitize@url \@href}%
\providecommand \@href[1]{\@@startlink{#1}\@@href}%
\providecommand \@@href[1]{\endgroup#1\@@endlink}%
\providecommand \@sanitize@url [0]{\catcode `\\12\catcode `\$12\catcode
  `\&12\catcode `\#12\catcode `\^12\catcode `\_12\catcode `\%12\relax}%
\providecommand \@@startlink[1]{}%
\providecommand \@@endlink[0]{}%
\providecommand \url  [0]{\begingroup\@sanitize@url \@url }%
\providecommand \@url [1]{\endgroup\@href {#1}{\urlprefix }}%
\providecommand \urlprefix  [0]{URL }%
\providecommand \Eprint [0]{\href }%
\providecommand \doibase [0]{http://dx.doi.org/}%
\providecommand \selectlanguage [0]{\@gobble}%
\providecommand \bibinfo  [0]{\@secondoftwo}%
\providecommand \bibfield  [0]{\@secondoftwo}%
\providecommand \translation [1]{[#1]}%
\providecommand \BibitemOpen [0]{}%
\providecommand \bibitemStop [0]{}%
\providecommand \bibitemNoStop [0]{.\EOS\space}%
\providecommand \EOS [0]{\spacefactor3000\relax}%
\providecommand \BibitemShut  [1]{\csname bibitem#1\endcsname}%
\let\auto@bib@innerbib\@empty
\bibitem [{\citenamefont {Mohseni}\ \emph {et~al.}(2017)\citenamefont
  {Mohseni}, \citenamefont {Read}, \citenamefont {Neven}, \citenamefont
  {Boixo}, \citenamefont {Denchev}, \citenamefont {Babbush}, \citenamefont
  {Fowler}, \citenamefont {Smelyanskiy},\ and\ \citenamefont
  {Martinis}}]{Mohseni2017}%
  \BibitemOpen
  \bibfield  {author} {\bibinfo {author} {\bibfnamefont {Masoud}\ \bibnamefont
  {Mohseni}}, \bibinfo {author} {\bibfnamefont {Peter}\ \bibnamefont {Read}},
  \bibinfo {author} {\bibfnamefont {Hartmut}\ \bibnamefont {Neven}}, \bibinfo
  {author} {\bibfnamefont {Sergio}\ \bibnamefont {Boixo}}, \bibinfo {author}
  {\bibfnamefont {Vasil}\ \bibnamefont {Denchev}}, \bibinfo {author}
  {\bibfnamefont {Ryan}\ \bibnamefont {Babbush}}, \bibinfo {author}
  {\bibfnamefont {Austin}\ \bibnamefont {Fowler}}, \bibinfo {author}
  {\bibfnamefont {Vadim}\ \bibnamefont {Smelyanskiy}}, \ and\ \bibinfo {author}
  {\bibfnamefont {John}\ \bibnamefont {Martinis}},\ }\bibfield  {title}
  {\enquote {\bibinfo {title} {Commercialize quantum technologies in five
  years},}\ }\href@noop {} {\bibfield  {journal} {\bibinfo  {journal} {Nature}\
  }\textbf {\bibinfo {volume} {543}},\ \bibinfo {pages} {171--174} (\bibinfo
  {year} {2017})}\BibitemShut {NoStop}%
\bibitem [{\citenamefont {Krizhevsky}\ \emph {et~al.}(2012)\citenamefont
  {Krizhevsky}, \citenamefont {Sutskever},\ and\ \citenamefont
  {Hinton}}]{krizhevsky2012imagenet}%
  \BibitemOpen
  \bibfield  {author} {\bibinfo {author} {\bibfnamefont {Alex}\ \bibnamefont
  {Krizhevsky}}, \bibinfo {author} {\bibfnamefont {Ilya}\ \bibnamefont
  {Sutskever}}, \ and\ \bibinfo {author} {\bibfnamefont {Geoffrey~E}\
  \bibnamefont {Hinton}},\ }\bibfield  {title} {\enquote {\bibinfo {title}
  {Imagenet classification with deep convolutional neural networks},}\ }in\
  \href@noop {} {\emph {\bibinfo {booktitle} {Advances in neural information
  processing systems}}}\ (\bibinfo {year} {2012})\ pp.\ \bibinfo {pages}
  {1097--1105}\BibitemShut {NoStop}%
\bibitem [{\citenamefont {Hinton}\ \emph {et~al.}(2012)\citenamefont {Hinton},
  \citenamefont {Deng}, \citenamefont {Yu}, \citenamefont {Dahl}, \citenamefont
  {Mohamed}, \citenamefont {Jaitly}, \citenamefont {Senior}, \citenamefont
  {Vanhoucke}, \citenamefont {Nguyen}, \citenamefont {Sainath} \emph
  {et~al.}}]{hinton2012deep}%
  \BibitemOpen
  \bibfield  {author} {\bibinfo {author} {\bibfnamefont {Geoffrey}\
  \bibnamefont {Hinton}}, \bibinfo {author} {\bibfnamefont {Li}~\bibnamefont
  {Deng}}, \bibinfo {author} {\bibfnamefont {Dong}\ \bibnamefont {Yu}},
  \bibinfo {author} {\bibfnamefont {George~E}\ \bibnamefont {Dahl}}, \bibinfo
  {author} {\bibfnamefont {Abdel-rahman}\ \bibnamefont {Mohamed}}, \bibinfo
  {author} {\bibfnamefont {Navdeep}\ \bibnamefont {Jaitly}}, \bibinfo {author}
  {\bibfnamefont {Andrew}\ \bibnamefont {Senior}}, \bibinfo {author}
  {\bibfnamefont {Vincent}\ \bibnamefont {Vanhoucke}}, \bibinfo {author}
  {\bibfnamefont {Patrick}\ \bibnamefont {Nguyen}}, \bibinfo {author}
  {\bibfnamefont {Tara~N}\ \bibnamefont {Sainath}},  \emph {et~al.},\
  }\bibfield  {title} {\enquote {\bibinfo {title} {Deep neural networks for
  acoustic modeling in speech recognition: The shared views of four research
  groups},}\ }\href@noop {} {\bibfield  {journal} {\bibinfo  {journal} {IEEE
  Signal Processing Magazine}\ }\textbf {\bibinfo {volume} {29}},\ \bibinfo
  {pages} {82--97} (\bibinfo {year} {2012})}\BibitemShut {NoStop}%
\bibitem [{\citenamefont {Levinson}\ \emph {et~al.}(2011)\citenamefont
  {Levinson}, \citenamefont {Askeland}, \citenamefont {Becker}, \citenamefont
  {Dolson}, \citenamefont {Held}, \citenamefont {Kammel}, \citenamefont
  {Kolter}, \citenamefont {Langer}, \citenamefont {Pink}, \citenamefont {Pratt}
  \emph {et~al.}}]{levinson2011towards}%
  \BibitemOpen
  \bibfield  {author} {\bibinfo {author} {\bibfnamefont {Jesse}\ \bibnamefont
  {Levinson}}, \bibinfo {author} {\bibfnamefont {Jake}\ \bibnamefont
  {Askeland}}, \bibinfo {author} {\bibfnamefont {Jan}\ \bibnamefont {Becker}},
  \bibinfo {author} {\bibfnamefont {Jennifer}\ \bibnamefont {Dolson}}, \bibinfo
  {author} {\bibfnamefont {David}\ \bibnamefont {Held}}, \bibinfo {author}
  {\bibfnamefont {Soeren}\ \bibnamefont {Kammel}}, \bibinfo {author}
  {\bibfnamefont {J~Zico}\ \bibnamefont {Kolter}}, \bibinfo {author}
  {\bibfnamefont {Dirk}\ \bibnamefont {Langer}}, \bibinfo {author}
  {\bibfnamefont {Oliver}\ \bibnamefont {Pink}}, \bibinfo {author}
  {\bibfnamefont {Vaughan}\ \bibnamefont {Pratt}},  \emph {et~al.},\ }\bibfield
   {title} {\enquote {\bibinfo {title} {Towards fully autonomous driving:
  Systems and algorithms},}\ }in\ \href@noop {} {\emph {\bibinfo {booktitle}
  {Intelligent Vehicles Symposium (IV), 2011 IEEE}}}\ (\bibinfo {organization}
  {IEEE},\ \bibinfo {year} {2011})\ pp.\ \bibinfo {pages}
  {163--168}\BibitemShut {NoStop}%
\bibitem [{\citenamefont {Esteva}\ \emph {et~al.}(2017)\citenamefont {Esteva},
  \citenamefont {Kuprel}, \citenamefont {Novoa}, \citenamefont {Ko},
  \citenamefont {Swetter}, \citenamefont {Blau},\ and\ \citenamefont
  {Thrun}}]{esteva2017dermatologist}%
  \BibitemOpen
  \bibfield  {author} {\bibinfo {author} {\bibfnamefont {Andre}\ \bibnamefont
  {Esteva}}, \bibinfo {author} {\bibfnamefont {Brett}\ \bibnamefont {Kuprel}},
  \bibinfo {author} {\bibfnamefont {Roberto~A}\ \bibnamefont {Novoa}}, \bibinfo
  {author} {\bibfnamefont {Justin}\ \bibnamefont {Ko}}, \bibinfo {author}
  {\bibfnamefont {Susan~M}\ \bibnamefont {Swetter}}, \bibinfo {author}
  {\bibfnamefont {Helen~M}\ \bibnamefont {Blau}}, \ and\ \bibinfo {author}
  {\bibfnamefont {Sebastian}\ \bibnamefont {Thrun}},\ }\bibfield  {title}
  {\enquote {\bibinfo {title} {Dermatologist-level classification of skin
  cancer with deep neural networks},}\ }\href@noop {} {\bibfield  {journal}
  {\bibinfo  {journal} {Nature}\ }\textbf {\bibinfo {volume} {542}},\ \bibinfo
  {pages} {115--118} (\bibinfo {year} {2017})}\BibitemShut {NoStop}%
\bibitem [{\citenamefont {Ching}\ \emph {et~al.}(2017)\citenamefont {Ching},
  \citenamefont {Himmelstein}, \citenamefont {Beaulieu-Jones}, \citenamefont
  {Kalinin}, \citenamefont {Do}, \citenamefont {Way}, \citenamefont {Ferrero},
  \citenamefont {Agapow}, \citenamefont {Xie}, \citenamefont {Rosen},
  \citenamefont {Lengerich}, \citenamefont {Israeli}, \citenamefont
  {Lanchantin}, \citenamefont {Woloszynek}, \citenamefont {Carpenter},
  \citenamefont {Shrikumar}, \citenamefont {Xu}, \citenamefont {Cofer},
  \citenamefont {Harris}, \citenamefont {DeCaprio}, \citenamefont {Qi},
  \citenamefont {Kundaje}, \citenamefont {Peng}, \citenamefont {Wiley},
  \citenamefont {Segler}, \citenamefont {Gitter},\ and\ \citenamefont
  {Greene}}]{Ching142760}%
  \BibitemOpen
  \bibfield  {author} {\bibinfo {author} {\bibfnamefont {Travers}\ \bibnamefont
  {Ching}}, \bibinfo {author} {\bibfnamefont {Daniel~S.}\ \bibnamefont
  {Himmelstein}}, \bibinfo {author} {\bibfnamefont {Brett~K.}\ \bibnamefont
  {Beaulieu-Jones}}, \bibinfo {author} {\bibfnamefont {Alexandr~A.}\
  \bibnamefont {Kalinin}}, \bibinfo {author} {\bibfnamefont {Brian~T.}\
  \bibnamefont {Do}}, \bibinfo {author} {\bibfnamefont {Gregory~P.}\
  \bibnamefont {Way}}, \bibinfo {author} {\bibfnamefont {Enrico}\ \bibnamefont
  {Ferrero}}, \bibinfo {author} {\bibfnamefont {Paul-Michael}\ \bibnamefont
  {Agapow}}, \bibinfo {author} {\bibfnamefont {Wei}\ \bibnamefont {Xie}},
  \bibinfo {author} {\bibfnamefont {Gail~L.}\ \bibnamefont {Rosen}}, \bibinfo
  {author} {\bibfnamefont {Benjamin~J.}\ \bibnamefont {Lengerich}}, \bibinfo
  {author} {\bibfnamefont {Johnny}\ \bibnamefont {Israeli}}, \bibinfo {author}
  {\bibfnamefont {Jack}\ \bibnamefont {Lanchantin}}, \bibinfo {author}
  {\bibfnamefont {Stephen}\ \bibnamefont {Woloszynek}}, \bibinfo {author}
  {\bibfnamefont {Anne~E.}\ \bibnamefont {Carpenter}}, \bibinfo {author}
  {\bibfnamefont {Avanti}\ \bibnamefont {Shrikumar}}, \bibinfo {author}
  {\bibfnamefont {Jinbo}\ \bibnamefont {Xu}}, \bibinfo {author} {\bibfnamefont
  {Evan~M.}\ \bibnamefont {Cofer}}, \bibinfo {author} {\bibfnamefont
  {David~J.}\ \bibnamefont {Harris}}, \bibinfo {author} {\bibfnamefont {Dave}\
  \bibnamefont {DeCaprio}}, \bibinfo {author} {\bibfnamefont {Yanjun}\
  \bibnamefont {Qi}}, \bibinfo {author} {\bibfnamefont {Anshul}\ \bibnamefont
  {Kundaje}}, \bibinfo {author} {\bibfnamefont {Yifan}\ \bibnamefont {Peng}},
  \bibinfo {author} {\bibfnamefont {Laura~K.}\ \bibnamefont {Wiley}}, \bibinfo
  {author} {\bibfnamefont {Marwin H.~S.}\ \bibnamefont {Segler}}, \bibinfo
  {author} {\bibfnamefont {Anthony}\ \bibnamefont {Gitter}}, \ and\ \bibinfo
  {author} {\bibfnamefont {Casey~S.}\ \bibnamefont {Greene}},\ }\bibfield
  {title} {\enquote {\bibinfo {title} {Opportunities and obstacles for deep
  learning in biology and medicine},}\ }\href {\doibase 10.1101/142760}
  {\bibfield  {journal} {\bibinfo  {journal} {bioRxiv}\ } (\bibinfo {year}
  {2017}),\ 10.1101/142760}\BibitemShut {NoStop}%
\bibitem [{\citenamefont {Mnih}\ \emph {et~al.}(2015)\citenamefont {Mnih},
  \citenamefont {Kavukcuoglu}, \citenamefont {Silver}, \citenamefont {Rusu},
  \citenamefont {Veness}, \citenamefont {Bellemare}, \citenamefont {Graves},
  \citenamefont {Riedmiller}, \citenamefont {Fidjeland}, \citenamefont
  {Ostrovski} \emph {et~al.}}]{mnih2015human}%
  \BibitemOpen
  \bibfield  {author} {\bibinfo {author} {\bibfnamefont {Volodymyr}\
  \bibnamefont {Mnih}}, \bibinfo {author} {\bibfnamefont {Koray}\ \bibnamefont
  {Kavukcuoglu}}, \bibinfo {author} {\bibfnamefont {David}\ \bibnamefont
  {Silver}}, \bibinfo {author} {\bibfnamefont {Andrei~A}\ \bibnamefont {Rusu}},
  \bibinfo {author} {\bibfnamefont {Joel}\ \bibnamefont {Veness}}, \bibinfo
  {author} {\bibfnamefont {Marc~G}\ \bibnamefont {Bellemare}}, \bibinfo
  {author} {\bibfnamefont {Alex}\ \bibnamefont {Graves}}, \bibinfo {author}
  {\bibfnamefont {Martin}\ \bibnamefont {Riedmiller}}, \bibinfo {author}
  {\bibfnamefont {Andreas~K}\ \bibnamefont {Fidjeland}}, \bibinfo {author}
  {\bibfnamefont {Georg}\ \bibnamefont {Ostrovski}},  \emph {et~al.},\
  }\bibfield  {title} {\enquote {\bibinfo {title} {Human-level control through
  deep reinforcement learning},}\ }\href@noop {} {\bibfield  {journal}
  {\bibinfo  {journal} {Nature}\ }\textbf {\bibinfo {volume} {518}},\ \bibinfo
  {pages} {529--533} (\bibinfo {year} {2015})}\BibitemShut {NoStop}%
\bibitem [{\citenamefont {Neven}\ \emph {et~al.}(2009)\citenamefont {Neven},
  \citenamefont {Denchev}, \citenamefont {Drew-Brook}, \citenamefont {Zhang},
  \citenamefont {Macready},\ and\ \citenamefont {Rose}}]{neven2009nips}%
  \BibitemOpen
  \bibfield  {author} {\bibinfo {author} {\bibfnamefont {Harmut}\ \bibnamefont
  {Neven}}, \bibinfo {author} {\bibfnamefont {Vasil~S}\ \bibnamefont
  {Denchev}}, \bibinfo {author} {\bibfnamefont {Marshall}\ \bibnamefont
  {Drew-Brook}}, \bibinfo {author} {\bibfnamefont {Jiayong}\ \bibnamefont
  {Zhang}}, \bibinfo {author} {\bibfnamefont {William~G}\ \bibnamefont
  {Macready}}, \ and\ \bibinfo {author} {\bibfnamefont {Geordie}\ \bibnamefont
  {Rose}},\ }\bibfield  {title} {\enquote {\bibinfo {title} {{Binary
  classification using hardware implementation of quantum annealing}},}\ }in\
  \href@noop {} {\emph {\bibinfo {booktitle} {{Demonstrations at NIPS-09, 24th
  Annual Conference on Neural Information Processing Systems}}}}\ (\bibinfo
  {year} {2009})\ pp.\ \bibinfo {pages} {1--17}\BibitemShut {NoStop}%
\bibitem [{\citenamefont {Bian}\ \emph {et~al.}(2010)\citenamefont {Bian},
  \citenamefont {Chudak}, \citenamefont {Macready},\ and\ \citenamefont
  {Rose}}]{bian2010ising}%
  \BibitemOpen
  \bibfield  {author} {\bibinfo {author} {\bibfnamefont {Zhengbing}\
  \bibnamefont {Bian}}, \bibinfo {author} {\bibfnamefont {Fabian}\ \bibnamefont
  {Chudak}}, \bibinfo {author} {\bibfnamefont {William~G}\ \bibnamefont
  {Macready}}, \ and\ \bibinfo {author} {\bibfnamefont {Geordie}\ \bibnamefont
  {Rose}},\ }\href@noop {} {\emph {\bibinfo {title} {{The Ising model: teaching
  an old problem new tricks}}}},\ \bibinfo {type} {Tech. Rep.}\ (\bibinfo
  {institution} {D-Wave Systems},\ \bibinfo {year} {2010})\BibitemShut
  {NoStop}%
\bibitem [{\citenamefont {Denil}\ and\ \citenamefont
  {De~Freitas}(2011)}]{Denil-2011}%
  \BibitemOpen
  \bibfield  {author} {\bibinfo {author} {\bibfnamefont {Misha}\ \bibnamefont
  {Denil}}\ and\ \bibinfo {author} {\bibfnamefont {Nando}\ \bibnamefont
  {De~Freitas}},\ }\bibfield  {title} {\enquote {\bibinfo {title} {Toward the
  implementation of a quantum {RBM}},}\ }\href@noop {} {\bibfield  {journal}
  {\bibinfo  {journal} {NIPS Deep Learning and Unsupervised Feature Learning
  Workshop}\ } (\bibinfo {year} {2011})}\BibitemShut {NoStop}%
\bibitem [{\citenamefont {Wiebe}\ \emph {et~al.}(2012)\citenamefont {Wiebe},
  \citenamefont {Braun},\ and\ \citenamefont {Lloyd}}]{wiebe2012quantum}%
  \BibitemOpen
  \bibfield  {author} {\bibinfo {author} {\bibfnamefont {Nathan}\ \bibnamefont
  {Wiebe}}, \bibinfo {author} {\bibfnamefont {Daniel}\ \bibnamefont {Braun}}, \
  and\ \bibinfo {author} {\bibfnamefont {Seth}\ \bibnamefont {Lloyd}},\
  }\bibfield  {title} {\enquote {\bibinfo {title} {Quantum algorithm for data
  fitting},}\ }\href@noop {} {\bibfield  {journal} {\bibinfo  {journal}
  {Physical review letters}\ }\textbf {\bibinfo {volume} {109}},\ \bibinfo
  {pages} {050505} (\bibinfo {year} {2012})}\BibitemShut {NoStop}%
\bibitem [{\citenamefont {Pudenz}\ and\ \citenamefont
  {Lidar}(2013)}]{Pudenz-QIP-2013}%
  \BibitemOpen
  \bibfield  {author} {\bibinfo {author} {\bibfnamefont {Kristen~L.}\
  \bibnamefont {Pudenz}}\ and\ \bibinfo {author} {\bibfnamefont {Daniel~A.}\
  \bibnamefont {Lidar}},\ }\bibfield  {title} {\enquote {\bibinfo {title}
  {Quantum adiabatic machine learning},}\ }\href {\doibase
  10.1007/s11128-012-0506-4} {\bibfield  {journal} {\bibinfo  {journal}
  {Quantum Information Processing}\ }\textbf {\bibinfo {volume} {12}},\
  \bibinfo {pages} {2027--2070} (\bibinfo {year} {2013})}\BibitemShut {NoStop}%
\bibitem [{\citenamefont {Lloyd}\ \emph {et~al.}(2013)\citenamefont {Lloyd},
  \citenamefont {Mohseni},\ and\ \citenamefont
  {Rebentrost}}]{Lloyd-arXiv-2013}%
  \BibitemOpen
  \bibfield  {author} {\bibinfo {author} {\bibfnamefont {Seth}\ \bibnamefont
  {Lloyd}}, \bibinfo {author} {\bibfnamefont {Masoud}\ \bibnamefont {Mohseni}},
  \ and\ \bibinfo {author} {\bibfnamefont {Patrick}\ \bibnamefont
  {Rebentrost}},\ }\bibfield  {title} {\enquote {\bibinfo {title} {Quantum
  algorithms for supervised and unsupervised machine learning},}\ }\href@noop
  {} {\bibfield  {journal} {\bibinfo  {journal} {arXiv:1307.0411}\ } (\bibinfo
  {year} {2013})}\BibitemShut {NoStop}%
\bibitem [{\citenamefont {Rebentrost}\ \emph {et~al.}(2014)\citenamefont
  {Rebentrost}, \citenamefont {Mohseni},\ and\ \citenamefont
  {Lloyd}}]{Rebentrost-PRL-2014}%
  \BibitemOpen
  \bibfield  {author} {\bibinfo {author} {\bibfnamefont {Patrick}\ \bibnamefont
  {Rebentrost}}, \bibinfo {author} {\bibfnamefont {Masoud}\ \bibnamefont
  {Mohseni}}, \ and\ \bibinfo {author} {\bibfnamefont {Seth}\ \bibnamefont
  {Lloyd}},\ }\bibfield  {title} {\enquote {\bibinfo {title} {Quantum support
  vector machine for big data classification},}\ }\href {\doibase
  10.1103/PhysRevLett.113.130503} {\bibfield  {journal} {\bibinfo  {journal}
  {Phys. Rev. Lett.}\ }\textbf {\bibinfo {volume} {113}},\ \bibinfo {pages}
  {130503} (\bibinfo {year} {2014})}\BibitemShut {NoStop}%
\bibitem [{\citenamefont {Wang}(2017)}]{wang2017quantum}%
  \BibitemOpen
  \bibfield  {author} {\bibinfo {author} {\bibfnamefont {Guoming}\ \bibnamefont
  {Wang}},\ }\bibfield  {title} {\enquote {\bibinfo {title} {Quantum algorithm
  for linear regression},}\ }\href@noop {} {\bibfield  {journal} {\bibinfo
  {journal} {Physical Review A}\ }\textbf {\bibinfo {volume} {96}},\ \bibinfo
  {pages} {012335} (\bibinfo {year} {2017})}\BibitemShut {NoStop}%
\bibitem [{\citenamefont {{Zhao}}\ \emph {et~al.}(2015)\citenamefont {{Zhao}},
  \citenamefont {{Fitzsimons}},\ and\ \citenamefont
  {{Fitzsimons}}}]{2015arXiv151203929Z}%
  \BibitemOpen
  \bibfield  {author} {\bibinfo {author} {\bibfnamefont {Z.}~\bibnamefont
  {{Zhao}}}, \bibinfo {author} {\bibfnamefont {J.~K.}\ \bibnamefont
  {{Fitzsimons}}}, \ and\ \bibinfo {author} {\bibfnamefont {J.~F.}\
  \bibnamefont {{Fitzsimons}}},\ }\bibfield  {title} {\enquote {\bibinfo
  {title} {{Quantum assisted Gaussian process regression}},}\ }\href@noop {}
  {\bibfield  {journal} {\bibinfo  {journal} {ArXiv e-prints}\ } (\bibinfo
  {year} {2015})},\ \Eprint {http://arxiv.org/abs/1512.03929} {arXiv:1512.03929
  [quant-ph]} \BibitemShut {NoStop}%
\bibitem [{\citenamefont {Lloyd}\ \emph {et~al.}(2014)\citenamefont {Lloyd},
  \citenamefont {Mohseni},\ and\ \citenamefont
  {Rebentrost}}]{Lloyd-NatPhys-2014}%
  \BibitemOpen
  \bibfield  {author} {\bibinfo {author} {\bibfnamefont {Seth}\ \bibnamefont
  {Lloyd}}, \bibinfo {author} {\bibfnamefont {Masoud}\ \bibnamefont {Mohseni}},
  \ and\ \bibinfo {author} {\bibfnamefont {Patrick}\ \bibnamefont
  {Rebentrost}},\ }\bibfield  {title} {\enquote {\bibinfo {title} {Quantum
  principal component analysis},}\ }\href@noop {} {\bibfield  {journal}
  {\bibinfo  {journal} {Nature Physics}\ }\textbf {\bibinfo {volume} {10}},\
  \bibinfo {pages} {631--633} (\bibinfo {year} {2014})}\BibitemShut {NoStop}%
\bibitem [{\citenamefont {Schuld}\ \emph {et~al.}(2016)\citenamefont {Schuld},
  \citenamefont {Sinayskiy},\ and\ \citenamefont
  {Petruccione}}]{schuld2016prediction}%
  \BibitemOpen
  \bibfield  {author} {\bibinfo {author} {\bibfnamefont {Maria}\ \bibnamefont
  {Schuld}}, \bibinfo {author} {\bibfnamefont {Ilya}\ \bibnamefont
  {Sinayskiy}}, \ and\ \bibinfo {author} {\bibfnamefont {Francesco}\
  \bibnamefont {Petruccione}},\ }\bibfield  {title} {\enquote {\bibinfo {title}
  {Prediction by linear regression on a quantum computer},}\ }\href@noop {}
  {\bibfield  {journal} {\bibinfo  {journal} {Physical Review A}\ }\textbf
  {\bibinfo {volume} {94}},\ \bibinfo {pages} {022342} (\bibinfo {year}
  {2016})}\BibitemShut {NoStop}%
\bibitem [{\citenamefont {Nathan~Wiebe}(2015)}]{Wiebe-arXiv-2015}%
  \BibitemOpen
  \bibfield  {author} {\bibinfo {author} {\bibfnamefont {Krysta M.~Svore}\
  \bibnamefont {Nathan~Wiebe}, \bibfnamefont {Ashish~Kapoor}},\ }\bibfield
  {title} {\enquote {\bibinfo {title} {Quantum deep learning},}\ }\href@noop {}
  {\bibfield  {journal} {\bibinfo  {journal} {arXiv:1412.3489}\ } (\bibinfo
  {year} {2015})}\BibitemShut {NoStop}%
\bibitem [{\citenamefont {Benedetti}\ \emph {et~al.}(2016)\citenamefont
  {Benedetti}, \citenamefont {Realpe-G\'omez}, \citenamefont {Biswas},\ and\
  \citenamefont {Perdomo-Ortiz}}]{Benedetti-2016}%
  \BibitemOpen
  \bibfield  {author} {\bibinfo {author} {\bibfnamefont {Marcello}\
  \bibnamefont {Benedetti}}, \bibinfo {author} {\bibfnamefont {John}\
  \bibnamefont {Realpe-G\'omez}}, \bibinfo {author} {\bibfnamefont {Rupak}\
  \bibnamefont {Biswas}}, \ and\ \bibinfo {author} {\bibfnamefont {Alejandro}\
  \bibnamefont {Perdomo-Ortiz}},\ }\bibfield  {title} {\enquote {\bibinfo
  {title} {Estimation of effective temperatures in quantum annealers for
  sampling applications: A case study with possible applications in deep
  learning},}\ }\href {\doibase 10.1103/PhysRevA.94.022308} {\bibfield
  {journal} {\bibinfo  {journal} {Phys. Rev. A}\ }\textbf {\bibinfo {volume}
  {94}},\ \bibinfo {pages} {022308} (\bibinfo {year} {2016})}\BibitemShut
  {NoStop}%
\bibitem [{\citenamefont {Benedetti}\ \emph
  {et~al.}(2017{\natexlab{a}})\citenamefont {Benedetti}, \citenamefont
  {Realpe-G\'omez}, \citenamefont {Biswas},\ and\ \citenamefont
  {Perdomo-Ortiz}}]{Benedetti2017}%
  \BibitemOpen
  \bibfield  {author} {\bibinfo {author} {\bibfnamefont {Marcello}\
  \bibnamefont {Benedetti}}, \bibinfo {author} {\bibfnamefont {John}\
  \bibnamefont {Realpe-G\'omez}}, \bibinfo {author} {\bibfnamefont {Rupak}\
  \bibnamefont {Biswas}}, \ and\ \bibinfo {author} {\bibfnamefont {Alejandro}\
  \bibnamefont {Perdomo-Ortiz}},\ }\bibfield  {title} {\enquote {\bibinfo
  {title} {Quantum-assisted learning of hardware-embedded probabilistic
  graphical models},}\ }\href {\doibase 10.1103/PhysRevX.7.041052} {\bibfield
  {journal} {\bibinfo  {journal} {Phys. Rev. X}\ }\textbf {\bibinfo {volume}
  {7}},\ \bibinfo {pages} {041052} (\bibinfo {year}
  {2017}{\natexlab{a}})}\BibitemShut {NoStop}%
\bibitem [{\citenamefont {Aaronson}(2015)}]{Aaronson-2015}%
  \BibitemOpen
  \bibfield  {author} {\bibinfo {author} {\bibfnamefont {Scott}\ \bibnamefont
  {Aaronson}},\ }\bibfield  {title} {\enquote {\bibinfo {title} {Read the fine
  print},}\ }\href@noop {} {\bibfield  {journal} {\bibinfo  {journal} {Nature
  Physics}\ }\textbf {\bibinfo {volume} {11}},\ \bibinfo {pages} {291--293}
  (\bibinfo {year} {2015})},\ \bibinfo {note} {commentary}\BibitemShut
  {NoStop}%
\bibitem [{\citenamefont {Adachi}\ and\ \citenamefont
  {Henderson}(2015)}]{Adachi-arXiv-2015}%
  \BibitemOpen
  \bibfield  {author} {\bibinfo {author} {\bibfnamefont {Steven~H.}\
  \bibnamefont {Adachi}}\ and\ \bibinfo {author} {\bibfnamefont {Maxwell~P.}\
  \bibnamefont {Henderson}},\ }\bibfield  {title} {\enquote {\bibinfo {title}
  {Application of quantum annealing to training of deep neural networks},}\
  }\href@noop {} {\bibfield  {journal} {\bibinfo  {journal} {arXiv:1510.06356}\
  } (\bibinfo {year} {2015})}\BibitemShut {NoStop}%
\bibitem [{\citenamefont {Chancellor}\ \emph {et~al.}(2016)\citenamefont
  {Chancellor}, \citenamefont {Szoke}, \citenamefont {Vinci}, \citenamefont
  {Aeppli},\ and\ \citenamefont {Warburton}}]{chancellor2016maximum}%
  \BibitemOpen
  \bibfield  {author} {\bibinfo {author} {\bibfnamefont {Nicholas}\
  \bibnamefont {Chancellor}}, \bibinfo {author} {\bibfnamefont {Szilard}\
  \bibnamefont {Szoke}}, \bibinfo {author} {\bibfnamefont {Walter}\
  \bibnamefont {Vinci}}, \bibinfo {author} {\bibfnamefont {Gabriel}\
  \bibnamefont {Aeppli}}, \ and\ \bibinfo {author} {\bibfnamefont {Paul~A}\
  \bibnamefont {Warburton}},\ }\bibfield  {title} {\enquote {\bibinfo {title}
  {Maximum-entropy inference with a programmable annealer},}\ }\href@noop {}
  {\bibfield  {journal} {\bibinfo  {journal} {Scientific reports}\ }\textbf
  {\bibinfo {volume} {6}} (\bibinfo {year} {2016})}\BibitemShut {NoStop}%
\bibitem [{\citenamefont {{Mohammad H. Amin and Evgeny Andriyash and Jason
  Rolfe and Bohdan Kulchytskyy and Roger Melko}}(2016)}]{Amin-arXiv-2016}%
  \BibitemOpen
  \bibfield  {author} {\bibinfo {author} {\bibnamefont {{Mohammad H. Amin and
  Evgeny Andriyash and Jason Rolfe and Bohdan Kulchytskyy and Roger Melko}}},\
  }\bibfield  {title} {\enquote {\bibinfo {title} {{Quantum Boltzmann
  Machine}},}\ }\href@noop {} {\bibfield  {journal} {\bibinfo  {journal}
  {arXiv:1601.02036}\ } (\bibinfo {year} {2016})}\BibitemShut {NoStop}%
\bibitem [{\citenamefont {Kieferov\'a}\ and\ \citenamefont
  {Wiebe}(2017)}]{kieferova2016tomography}%
  \BibitemOpen
  \bibfield  {author} {\bibinfo {author} {\bibfnamefont {M\'aria}\ \bibnamefont
  {Kieferov\'a}}\ and\ \bibinfo {author} {\bibfnamefont {Nathan}\ \bibnamefont
  {Wiebe}},\ }\bibfield  {title} {\enquote {\bibinfo {title} {Tomography and
  generative training with quantum boltzmann machines},}\ }\href {\doibase
  10.1103/PhysRevA.96.062327} {\bibfield  {journal} {\bibinfo  {journal} {Phys.
  Rev. A}\ }\textbf {\bibinfo {volume} {96}},\ \bibinfo {pages} {062327}
  (\bibinfo {year} {2017})}\BibitemShut {NoStop}%
\bibitem [{\citenamefont {Kerenidis}\ and\ \citenamefont
  {Prakash}(2016)}]{kerenidis2016quantum}%
  \BibitemOpen
  \bibfield  {author} {\bibinfo {author} {\bibfnamefont {Iordanis}\
  \bibnamefont {Kerenidis}}\ and\ \bibinfo {author} {\bibfnamefont {Anupam}\
  \bibnamefont {Prakash}},\ }\bibfield  {title} {\enquote {\bibinfo {title}
  {Quantum recommendation systems},}\ }\href@noop {} {\bibfield  {journal}
  {\bibinfo  {journal} {arXiv preprint arXiv:1603.08675}\ } (\bibinfo {year}
  {2016})}\BibitemShut {NoStop}%
\bibitem [{\citenamefont {Wittek}\ and\ \citenamefont
  {Gogolin}(2017)}]{wittek2017quantum}%
  \BibitemOpen
  \bibfield  {author} {\bibinfo {author} {\bibfnamefont {Peter}\ \bibnamefont
  {Wittek}}\ and\ \bibinfo {author} {\bibfnamefont {Christian}\ \bibnamefont
  {Gogolin}},\ }\bibfield  {title} {\enquote {\bibinfo {title} {Quantum
  enhanced inference in markov logic networks},}\ }\href@noop {} {\bibfield
  {journal} {\bibinfo  {journal} {Scientific Reports}\ }\textbf {\bibinfo
  {volume} {7}} (\bibinfo {year} {2017})}\BibitemShut {NoStop}%
\bibitem [{\citenamefont {Potok}\ \emph {et~al.}(2017)\citenamefont {Potok},
  \citenamefont {Schuman}, \citenamefont {Young}, \citenamefont {Patton},
  \citenamefont {Spedalieri}, \citenamefont {Liu}, \citenamefont {Yao},
  \citenamefont {Rose},\ and\ \citenamefont {Chakma}}]{Potok2017}%
  \BibitemOpen
  \bibfield  {author} {\bibinfo {author} {\bibfnamefont {Thomas~E.}\
  \bibnamefont {Potok}}, \bibinfo {author} {\bibfnamefont {Catherine}\
  \bibnamefont {Schuman}}, \bibinfo {author} {\bibfnamefont {Steven~R.}\
  \bibnamefont {Young}}, \bibinfo {author} {\bibfnamefont {Robert~M.}\
  \bibnamefont {Patton}}, \bibinfo {author} {\bibfnamefont {Federico}\
  \bibnamefont {Spedalieri}}, \bibinfo {author} {\bibfnamefont {Jeremy}\
  \bibnamefont {Liu}}, \bibinfo {author} {\bibfnamefont {Ke-Thia}\ \bibnamefont
  {Yao}}, \bibinfo {author} {\bibfnamefont {Garrett}\ \bibnamefont {Rose}}, \
  and\ \bibinfo {author} {\bibfnamefont {Gangotree}\ \bibnamefont {Chakma}},\
  }\bibfield  {title} {\enquote {\bibinfo {title} {A study of complex deep
  learning networks on high performance, neuromorphic, and quantum
  computers},}\ }\href@noop {} {\bibfield  {journal} {\bibinfo  {journal}
  {arXiv:1703.05364}\ } (\bibinfo {year} {2017})}\BibitemShut {NoStop}%
\bibitem [{\citenamefont {Schuld}\ \emph {et~al.}(2015)\citenamefont {Schuld},
  \citenamefont {Sinayskiy},\ and\ \citenamefont
  {Petruccione}}]{Schuld-QML-2015}%
  \BibitemOpen
  \bibfield  {author} {\bibinfo {author} {\bibfnamefont {Maria}\ \bibnamefont
  {Schuld}}, \bibinfo {author} {\bibfnamefont {Ilya}\ \bibnamefont
  {Sinayskiy}}, \ and\ \bibinfo {author} {\bibfnamefont {Francesco}\
  \bibnamefont {Petruccione}},\ }\bibfield  {title} {\enquote {\bibinfo {title}
  {An introduction to quantum machine learning},}\ }\href@noop {} {\bibfield
  {journal} {\bibinfo  {journal} {Contemporary Physics}\ }\textbf {\bibinfo
  {volume} {56}},\ \bibinfo {pages} {172--185} (\bibinfo {year}
  {2015})}\BibitemShut {NoStop}%
\bibitem [{\citenamefont {Romero}\ \emph {et~al.}(2017)\citenamefont {Romero},
  \citenamefont {Olson},\ and\ \citenamefont {Aspuru-Guzik}}]{Romero2017}%
  \BibitemOpen
  \bibfield  {author} {\bibinfo {author} {\bibfnamefont {Jonathan}\
  \bibnamefont {Romero}}, \bibinfo {author} {\bibfnamefont {Jonathan~P}\
  \bibnamefont {Olson}}, \ and\ \bibinfo {author} {\bibfnamefont {Alan}\
  \bibnamefont {Aspuru-Guzik}},\ }\bibfield  {title} {\enquote {\bibinfo
  {title} {Quantum autoencoders for efficient compression of quantum data},}\
  }\href@noop {} {\bibfield  {journal} {\bibinfo  {journal} {Quantum Sci.
  Technol.}\ }\textbf {\bibinfo {volume} {2}},\ \bibinfo {pages} {045001}
  (\bibinfo {year} {2017})}\BibitemShut {NoStop}%
\bibitem [{\citenamefont {Adcock}\ \emph {et~al.}(2015)\citenamefont {Adcock},
  \citenamefont {Allen}, \citenamefont {Day}, \citenamefont {Frick},
  \citenamefont {Hinchliff}, \citenamefont {Johnson}, \citenamefont
  {Morley-Short}, \citenamefont {Pallister}, \citenamefont {Price},\ and\
  \citenamefont {Stanisic}}]{adcock2015advances}%
  \BibitemOpen
  \bibfield  {author} {\bibinfo {author} {\bibfnamefont {Jeremy}\ \bibnamefont
  {Adcock}}, \bibinfo {author} {\bibfnamefont {Euan}\ \bibnamefont {Allen}},
  \bibinfo {author} {\bibfnamefont {Matthew}\ \bibnamefont {Day}}, \bibinfo
  {author} {\bibfnamefont {Stefan}\ \bibnamefont {Frick}}, \bibinfo {author}
  {\bibfnamefont {Janna}\ \bibnamefont {Hinchliff}}, \bibinfo {author}
  {\bibfnamefont {Mack}\ \bibnamefont {Johnson}}, \bibinfo {author}
  {\bibfnamefont {Sam}\ \bibnamefont {Morley-Short}}, \bibinfo {author}
  {\bibfnamefont {Sam}\ \bibnamefont {Pallister}}, \bibinfo {author}
  {\bibfnamefont {Alasdair}\ \bibnamefont {Price}}, \ and\ \bibinfo {author}
  {\bibfnamefont {Stasja}\ \bibnamefont {Stanisic}},\ }\bibfield  {title}
  {\enquote {\bibinfo {title} {Advances in quantum machine learning},}\
  }\href@noop {} {\bibfield  {journal} {\bibinfo  {journal} {arXiv preprint
  arXiv:1512.02900}\ } (\bibinfo {year} {2015})}\BibitemShut {NoStop}%
\bibitem [{\citenamefont {Biamonte}\ \emph {et~al.}(2016)\citenamefont
  {Biamonte}, \citenamefont {Wittek}, \citenamefont {Pancotti}, \citenamefont
  {Rebentrost}, \citenamefont {Wiebe},\ and\ \citenamefont
  {Lloyd}}]{biamonte2016quantum}%
  \BibitemOpen
  \bibfield  {author} {\bibinfo {author} {\bibfnamefont {Jacob}\ \bibnamefont
  {Biamonte}}, \bibinfo {author} {\bibfnamefont {Peter}\ \bibnamefont
  {Wittek}}, \bibinfo {author} {\bibfnamefont {Nicola}\ \bibnamefont
  {Pancotti}}, \bibinfo {author} {\bibfnamefont {Patrick}\ \bibnamefont
  {Rebentrost}}, \bibinfo {author} {\bibfnamefont {Nathan}\ \bibnamefont
  {Wiebe}}, \ and\ \bibinfo {author} {\bibfnamefont {Seth}\ \bibnamefont
  {Lloyd}},\ }\bibfield  {title} {\enquote {\bibinfo {title} {Quantum machine
  learning},}\ }\href@noop {} {\bibfield  {journal} {\bibinfo  {journal} {arXiv
  preprint arXiv:1611.09347}\ } (\bibinfo {year} {2016})}\BibitemShut {NoStop}%
\bibitem [{\citenamefont {Alvarez-Rodriguez}\ \emph {et~al.}(2016)\citenamefont
  {Alvarez-Rodriguez}, \citenamefont {Lamata}, \citenamefont
  {Escandell-Montero}, \citenamefont {Mart{\'\i}n-Guerrero},\ and\
  \citenamefont {Solano}}]{alvarez2016quantum}%
  \BibitemOpen
  \bibfield  {author} {\bibinfo {author} {\bibfnamefont {Unai}\ \bibnamefont
  {Alvarez-Rodriguez}}, \bibinfo {author} {\bibfnamefont {Lucas}\ \bibnamefont
  {Lamata}}, \bibinfo {author} {\bibfnamefont {Pablo}\ \bibnamefont
  {Escandell-Montero}}, \bibinfo {author} {\bibfnamefont {Jos{\'e}~D}\
  \bibnamefont {Mart{\'\i}n-Guerrero}}, \ and\ \bibinfo {author} {\bibfnamefont
  {Enrique}\ \bibnamefont {Solano}},\ }\bibfield  {title} {\enquote {\bibinfo
  {title} {Quantum machine learning without measurements},}\ }\href@noop {}
  {\bibfield  {journal} {\bibinfo  {journal} {arXiv preprint arXiv:1612.05535}\
  } (\bibinfo {year} {2016})}\BibitemShut {NoStop}%
\bibitem [{\citenamefont {Lamata}(2017)}]{lamata2017basic}%
  \BibitemOpen
  \bibfield  {author} {\bibinfo {author} {\bibfnamefont {Lucas}\ \bibnamefont
  {Lamata}},\ }\bibfield  {title} {\enquote {\bibinfo {title} {Basic protocols
  in quantum reinforcement learning with superconducting circuits},}\
  }\href@noop {} {\bibfield  {journal} {\bibinfo  {journal} {Scientific
  Reports}\ }\textbf {\bibinfo {volume} {7}} (\bibinfo {year}
  {2017})}\BibitemShut {NoStop}%
\bibitem [{\citenamefont {Schuld}\ \emph {et~al.}(2017)\citenamefont {Schuld},
  \citenamefont {Fingerhuth},\ and\ \citenamefont
  {Petruccione}}]{schuld2017quantum}%
  \BibitemOpen
  \bibfield  {author} {\bibinfo {author} {\bibfnamefont {Maria}\ \bibnamefont
  {Schuld}}, \bibinfo {author} {\bibfnamefont {Mark}\ \bibnamefont
  {Fingerhuth}}, \ and\ \bibinfo {author} {\bibfnamefont {Francesco}\
  \bibnamefont {Petruccione}},\ }\bibfield  {title} {\enquote {\bibinfo {title}
  {Quantum machine learning with small-scale devices: Implementing a
  distance-based classifier with a quantum interference circuit},}\ }\href@noop
  {} {\bibfield  {journal} {\bibinfo  {journal} {arXiv preprint
  arXiv:1703.10793}\ } (\bibinfo {year} {2017})}\BibitemShut {NoStop}%
\bibitem [{\citenamefont {{Ciliberto}}\ \emph {et~al.}(2017)\citenamefont
  {{Ciliberto}}, \citenamefont {{Herbster}}, \citenamefont {{Davide Ialongo}},
  \citenamefont {{Pontil}}, \citenamefont {{Rocchetto}}, \citenamefont
  {{Severini}},\ and\ \citenamefont {{Wossnig}}}]{2017arXiv170708561C}%
  \BibitemOpen
  \bibfield  {author} {\bibinfo {author} {\bibfnamefont {C.}~\bibnamefont
  {{Ciliberto}}}, \bibinfo {author} {\bibfnamefont {M.}~\bibnamefont
  {{Herbster}}}, \bibinfo {author} {\bibfnamefont {A.}~\bibnamefont {{Davide
  Ialongo}}}, \bibinfo {author} {\bibfnamefont {M.}~\bibnamefont {{Pontil}}},
  \bibinfo {author} {\bibfnamefont {A.}~\bibnamefont {{Rocchetto}}}, \bibinfo
  {author} {\bibfnamefont {S.}~\bibnamefont {{Severini}}}, \ and\ \bibinfo
  {author} {\bibfnamefont {L.}~\bibnamefont {{Wossnig}}},\ }\bibfield  {title}
  {\enquote {\bibinfo {title} {{Quantum machine learning: a classical
  perspective}},}\ }\href@noop {} {\bibfield  {journal} {\bibinfo  {journal}
  {ArXiv e-prints}\ } (\bibinfo {year} {2017})},\ \Eprint
  {http://arxiv.org/abs/1707.08561} {arXiv:1707.08561 [quant-ph]} \BibitemShut
  {NoStop}%
\bibitem [{\citenamefont {Benedetti}\ \emph
  {et~al.}(2017{\natexlab{b}})\citenamefont {Benedetti}, \citenamefont
  {Realpe-G\'omez},\ and\ \citenamefont {Perdomo-Ortiz}}]{Benedetti2017b}%
  \BibitemOpen
  \bibfield  {author} {\bibinfo {author} {\bibfnamefont {Marcello}\
  \bibnamefont {Benedetti}}, \bibinfo {author} {\bibfnamefont {John}\
  \bibnamefont {Realpe-G\'omez}}, \ and\ \bibinfo {author} {\bibfnamefont
  {Alejandro}\ \bibnamefont {Perdomo-Ortiz}},\ }\bibfield  {title} {\enquote
  {\bibinfo {title} {Quantum-assisted helmholtz machines: A quantum-classical
  deep learning framework for industrial datasets in near-term devices},}\
  }\href@noop {} {\bibfield  {journal} {\bibinfo  {journal} {arXiv:1708.09784}\
  } (\bibinfo {year} {2017}{\natexlab{b}})}\BibitemShut {NoStop}%
\bibitem [{\citenamefont {Benedetti}\ \emph {et~al.}(2018)\citenamefont
  {Benedetti}, \citenamefont {Garcia-Pintos}, \citenamefont {Nam},\ and\
  \citenamefont {Perdomo-Ortiz}}]{Benedetti2018}%
  \BibitemOpen
  \bibfield  {author} {\bibinfo {author} {\bibfnamefont {Marcello}\
  \bibnamefont {Benedetti}}, \bibinfo {author} {\bibfnamefont {Delfina}\
  \bibnamefont {Garcia-Pintos}}, \bibinfo {author} {\bibfnamefont {Yunseong}\
  \bibnamefont {Nam}}, \ and\ \bibinfo {author} {\bibfnamefont {Alejandro}\
  \bibnamefont {Perdomo-Ortiz}},\ }\bibfield  {title} {\enquote {\bibinfo
  {title} {A generative modeling approach for benchmarking and training shallow
  quantum circuits},}\ }\href@noop {} {\bibfield  {journal} {\bibinfo
  {journal} {arXiv:1801.07686}\ } (\bibinfo {year} {2018})}\BibitemShut
  {NoStop}%
\bibitem [{\citenamefont {Farhi}\ and\ \citenamefont
  {Neven}(2018)}]{Farhi2018}%
  \BibitemOpen
  \bibfield  {author} {\bibinfo {author} {\bibfnamefont {Edward}\ \bibnamefont
  {Farhi}}\ and\ \bibinfo {author} {\bibfnamefont {Hartmut}\ \bibnamefont
  {Neven}},\ }\bibfield  {title} {\enquote {\bibinfo {title} {Classification
  with quantum neural networks on near term processors},}\ }\href@noop {}
  {\bibfield  {journal} {\bibinfo  {journal} {arXiv:1802.06002}\ } (\bibinfo
  {year} {2018})}\BibitemShut {NoStop}%
\bibitem [{\citenamefont {Bengio}\ \emph
  {et~al.}(2013{\natexlab{a}})\citenamefont {Bengio}, \citenamefont
  {Courville},\ and\ \citenamefont {Vincent}}]{bengio2013representation}%
  \BibitemOpen
  \bibfield  {author} {\bibinfo {author} {\bibfnamefont {Yoshua}\ \bibnamefont
  {Bengio}}, \bibinfo {author} {\bibfnamefont {Aaron}\ \bibnamefont
  {Courville}}, \ and\ \bibinfo {author} {\bibfnamefont {Pascal}\ \bibnamefont
  {Vincent}},\ }\bibfield  {title} {\enquote {\bibinfo {title} {Representation
  learning: A review and new perspectives},}\ }\href@noop {} {\bibfield
  {journal} {\bibinfo  {journal} {IEEE transactions on pattern analysis and
  machine intelligence}\ }\textbf {\bibinfo {volume} {35}},\ \bibinfo {pages}
  {1798--1828} (\bibinfo {year} {2013}{\natexlab{a}})}\BibitemShut {NoStop}%
\bibitem [{\citenamefont {Erhan}\ \emph {et~al.}(2010)\citenamefont {Erhan},
  \citenamefont {Bengio}, \citenamefont {Courville}, \citenamefont {Manzagol},
  \citenamefont {Vincent},\ and\ \citenamefont {Bengio}}]{erhan2010does}%
  \BibitemOpen
  \bibfield  {author} {\bibinfo {author} {\bibfnamefont {Dumitru}\ \bibnamefont
  {Erhan}}, \bibinfo {author} {\bibfnamefont {Yoshua}\ \bibnamefont {Bengio}},
  \bibinfo {author} {\bibfnamefont {Aaron}\ \bibnamefont {Courville}}, \bibinfo
  {author} {\bibfnamefont {Pierre-Antoine}\ \bibnamefont {Manzagol}}, \bibinfo
  {author} {\bibfnamefont {Pascal}\ \bibnamefont {Vincent}}, \ and\ \bibinfo
  {author} {\bibfnamefont {Samy}\ \bibnamefont {Bengio}},\ }\bibfield  {title}
  {\enquote {\bibinfo {title} {Why does unsupervised pre-training help deep
  learning?}}\ }\href@noop {} {\bibfield  {journal} {\bibinfo  {journal}
  {Journal of Machine Learning Research}\ }\textbf {\bibinfo {volume} {11}},\
  \bibinfo {pages} {625--660} (\bibinfo {year} {2010})}\BibitemShut {NoStop}%
\bibitem [{\citenamefont {Goodfellow}(2016)}]{goodfellow2016nips}%
  \BibitemOpen
  \bibfield  {author} {\bibinfo {author} {\bibfnamefont {Ian}\ \bibnamefont
  {Goodfellow}},\ }\bibfield  {title} {\enquote {\bibinfo {title} {Nips 2016
  tutorial: Generative adversarial networks},}\ }\href@noop {} {\bibfield
  {journal} {\bibinfo  {journal} {arXiv:1701.00160}\ } (\bibinfo {year}
  {2016})}\BibitemShut {NoStop}%
\bibitem [{\citenamefont {Kingma}\ and\ \citenamefont
  {Welling}(2013)}]{kingma2013auto}%
  \BibitemOpen
  \bibfield  {author} {\bibinfo {author} {\bibfnamefont {Diederik~P}\
  \bibnamefont {Kingma}}\ and\ \bibinfo {author} {\bibfnamefont {Max}\
  \bibnamefont {Welling}},\ }\bibfield  {title} {\enquote {\bibinfo {title}
  {Auto-encoding variational bayes},}\ }\href@noop {} {\bibfield  {journal}
  {\bibinfo  {journal} {arXiv preprint arXiv:1312.6114}\ } (\bibinfo {year}
  {2013})}\BibitemShut {NoStop}%
\bibitem [{\citenamefont {Rezende}\ \emph {et~al.}(2014)\citenamefont
  {Rezende}, \citenamefont {Mohamed},\ and\ \citenamefont
  {Wierstra}}]{rezende2014stochastic}%
  \BibitemOpen
  \bibfield  {author} {\bibinfo {author} {\bibfnamefont {Danilo~Jimenez}\
  \bibnamefont {Rezende}}, \bibinfo {author} {\bibfnamefont {Shakir}\
  \bibnamefont {Mohamed}}, \ and\ \bibinfo {author} {\bibfnamefont {Daan}\
  \bibnamefont {Wierstra}},\ }\bibfield  {title} {\enquote {\bibinfo {title}
  {Stochastic backpropagation and approximate inference in deep generative
  models},}\ }\href@noop {} {\bibfield  {journal} {\bibinfo  {journal} {arXiv
  preprint arXiv:1401.4082}\ } (\bibinfo {year} {2014})}\BibitemShut {NoStop}%
\bibitem [{\citenamefont {Mnih}\ and\ \citenamefont
  {Gregor}(2014)}]{mnih2014neural}%
  \BibitemOpen
  \bibfield  {author} {\bibinfo {author} {\bibfnamefont {Andriy}\ \bibnamefont
  {Mnih}}\ and\ \bibinfo {author} {\bibfnamefont {Karol}\ \bibnamefont
  {Gregor}},\ }\bibfield  {title} {\enquote {\bibinfo {title} {Neural
  variational inference and learning in belief networks},}\ }\href@noop {}
  {\bibfield  {journal} {\bibinfo  {journal} {arXiv preprint arXiv:1402.0030}\
  } (\bibinfo {year} {2014})}\BibitemShut {NoStop}%
\bibitem [{\citenamefont {S{\o}nderby}\ \emph {et~al.}(2016)\citenamefont
  {S{\o}nderby}, \citenamefont {Raiko}, \citenamefont {Maal{\o}e},
  \citenamefont {S{\o}nderby},\ and\ \citenamefont
  {Winther}}]{sonderby2016ladder}%
  \BibitemOpen
  \bibfield  {author} {\bibinfo {author} {\bibfnamefont {Casper~Kaae}\
  \bibnamefont {S{\o}nderby}}, \bibinfo {author} {\bibfnamefont {Tapani}\
  \bibnamefont {Raiko}}, \bibinfo {author} {\bibfnamefont {Lars}\ \bibnamefont
  {Maal{\o}e}}, \bibinfo {author} {\bibfnamefont {S{\o}ren~Kaae}\ \bibnamefont
  {S{\o}nderby}}, \ and\ \bibinfo {author} {\bibfnamefont {Ole}\ \bibnamefont
  {Winther}},\ }\bibfield  {title} {\enquote {\bibinfo {title} {Ladder
  variational autoencoders},}\ }in\ \href@noop {} {\emph {\bibinfo {booktitle}
  {Advances in Neural Information Processing Systems}}}\ (\bibinfo {year}
  {2016})\ pp.\ \bibinfo {pages} {3738--3746}\BibitemShut {NoStop}%
\bibitem [{\citenamefont {Rolfe}(2016)}]{rolfe2016discrete}%
  \BibitemOpen
  \bibfield  {author} {\bibinfo {author} {\bibfnamefont {Jason~Tyler}\
  \bibnamefont {Rolfe}},\ }\bibfield  {title} {\enquote {\bibinfo {title}
  {Discrete variational autoencoders},}\ }\href@noop {} {\bibfield  {journal}
  {\bibinfo  {journal} {arXiv preprint arXiv:1609.02200}\ } (\bibinfo {year}
  {2016})}\BibitemShut {NoStop}%
\bibitem [{\citenamefont {Ackley}\ \emph {et~al.}(1985)\citenamefont {Ackley},
  \citenamefont {Hinton},\ and\ \citenamefont
  {Sejnowski}}]{ackley1985learning}%
  \BibitemOpen
  \bibfield  {author} {\bibinfo {author} {\bibfnamefont {David~H}\ \bibnamefont
  {Ackley}}, \bibinfo {author} {\bibfnamefont {Geoffrey~E}\ \bibnamefont
  {Hinton}}, \ and\ \bibinfo {author} {\bibfnamefont {Terrence~J}\ \bibnamefont
  {Sejnowski}},\ }\bibfield  {title} {\enquote {\bibinfo {title} {A learning
  algorithm for boltzmann machines},}\ }\href@noop {} {\bibfield  {journal}
  {\bibinfo  {journal} {Cognitive science}\ }\textbf {\bibinfo {volume} {9}},\
  \bibinfo {pages} {147--169} (\bibinfo {year} {1985})}\BibitemShut {NoStop}%
\bibitem [{\citenamefont {Hinton}\ \emph {et~al.}(2006)\citenamefont {Hinton},
  \citenamefont {Osindero},\ and\ \citenamefont {Teh}}]{hinton2006fast}%
  \BibitemOpen
  \bibfield  {author} {\bibinfo {author} {\bibfnamefont {Geoffrey~E}\
  \bibnamefont {Hinton}}, \bibinfo {author} {\bibfnamefont {Simon}\
  \bibnamefont {Osindero}}, \ and\ \bibinfo {author} {\bibfnamefont {Yee-Whye}\
  \bibnamefont {Teh}},\ }\bibfield  {title} {\enquote {\bibinfo {title} {A fast
  learning algorithm for deep belief nets},}\ }\href@noop {} {\bibfield
  {journal} {\bibinfo  {journal} {Neural computation}\ }\textbf {\bibinfo
  {volume} {18}},\ \bibinfo {pages} {1527--1554} (\bibinfo {year}
  {2006})}\BibitemShut {NoStop}%
\bibitem [{\citenamefont {Bornschein}\ and\ \citenamefont
  {Bengio}(2014)}]{bornschein2014reweighted}%
  \BibitemOpen
  \bibfield  {author} {\bibinfo {author} {\bibfnamefont {J{\"o}rg}\
  \bibnamefont {Bornschein}}\ and\ \bibinfo {author} {\bibfnamefont {Yoshua}\
  \bibnamefont {Bengio}},\ }\bibfield  {title} {\enquote {\bibinfo {title}
  {Reweighted wake-sleep},}\ }\href@noop {} {\bibfield  {journal} {\bibinfo
  {journal} {arXiv preprint arXiv:1406.2751}\ } (\bibinfo {year}
  {2014})}\BibitemShut {NoStop}%
\bibitem [{\citenamefont {Bornschein}\ \emph {et~al.}(2016)\citenamefont
  {Bornschein}, \citenamefont {Shabanian}, \citenamefont {Fischer},\ and\
  \citenamefont {Bengio}}]{bornschein2016bidirectional}%
  \BibitemOpen
  \bibfield  {author} {\bibinfo {author} {\bibfnamefont {Jorg}\ \bibnamefont
  {Bornschein}}, \bibinfo {author} {\bibfnamefont {Samira}\ \bibnamefont
  {Shabanian}}, \bibinfo {author} {\bibfnamefont {Asja}\ \bibnamefont
  {Fischer}}, \ and\ \bibinfo {author} {\bibfnamefont {Yoshua}\ \bibnamefont
  {Bengio}},\ }\bibfield  {title} {\enquote {\bibinfo {title} {Bidirectional
  helmholtz machines},}\ }in\ \href@noop {} {\emph {\bibinfo {booktitle}
  {International Conference on Machine Learning}}}\ (\bibinfo {year} {2016})\
  pp.\ \bibinfo {pages} {2511--2519}\BibitemShut {NoStop}%
\bibitem [{\citenamefont {Salakhutdinov}\ and\ \citenamefont
  {Hinton}(2009)}]{salakhutdinov2009deep}%
  \BibitemOpen
  \bibfield  {author} {\bibinfo {author} {\bibfnamefont {Ruslan}\ \bibnamefont
  {Salakhutdinov}}\ and\ \bibinfo {author} {\bibfnamefont {Geoffrey}\
  \bibnamefont {Hinton}},\ }\bibfield  {title} {\enquote {\bibinfo {title}
  {Deep boltzmann machines},}\ }in\ \href@noop {} {\emph {\bibinfo {booktitle}
  {Artificial Intelligence and Statistics}}}\ (\bibinfo {year} {2009})\ pp.\
  \bibinfo {pages} {448--455}\BibitemShut {NoStop}%
\bibitem [{\citenamefont {Bengio}\ \emph {et~al.}(2014)\citenamefont {Bengio},
  \citenamefont {Laufer}, \citenamefont {Alain},\ and\ \citenamefont
  {Yosinski}}]{bengio2014deep}%
  \BibitemOpen
  \bibfield  {author} {\bibinfo {author} {\bibfnamefont {Yoshua}\ \bibnamefont
  {Bengio}}, \bibinfo {author} {\bibfnamefont {Eric}\ \bibnamefont {Laufer}},
  \bibinfo {author} {\bibfnamefont {Guillaume}\ \bibnamefont {Alain}}, \ and\
  \bibinfo {author} {\bibfnamefont {Jason}\ \bibnamefont {Yosinski}},\
  }\bibfield  {title} {\enquote {\bibinfo {title} {Deep generative stochastic
  networks trainable by backprop},}\ }in\ \href@noop {} {\emph {\bibinfo
  {booktitle} {International Conference on Machine Learning}}}\ (\bibinfo
  {year} {2014})\ pp.\ \bibinfo {pages} {226--234}\BibitemShut {NoStop}%
\bibitem [{\citenamefont {Goodfellow}\ \emph {et~al.}(2014)\citenamefont
  {Goodfellow}, \citenamefont {Pouget-Abadie}, \citenamefont {Mirza},
  \citenamefont {Xu}, \citenamefont {Warde-Farley}, \citenamefont {Ozair},
  \citenamefont {Courville},\ and\ \citenamefont
  {Bengio}}]{goodfellow2014generative}%
  \BibitemOpen
  \bibfield  {author} {\bibinfo {author} {\bibfnamefont {Ian}\ \bibnamefont
  {Goodfellow}}, \bibinfo {author} {\bibfnamefont {Jean}\ \bibnamefont
  {Pouget-Abadie}}, \bibinfo {author} {\bibfnamefont {Mehdi}\ \bibnamefont
  {Mirza}}, \bibinfo {author} {\bibfnamefont {Bing}\ \bibnamefont {Xu}},
  \bibinfo {author} {\bibfnamefont {David}\ \bibnamefont {Warde-Farley}},
  \bibinfo {author} {\bibfnamefont {Sherjil}\ \bibnamefont {Ozair}}, \bibinfo
  {author} {\bibfnamefont {Aaron}\ \bibnamefont {Courville}}, \ and\ \bibinfo
  {author} {\bibfnamefont {Yoshua}\ \bibnamefont {Bengio}},\ }\bibfield
  {title} {\enquote {\bibinfo {title} {Generative adversarial nets},}\ }in\
  \href@noop {} {\emph {\bibinfo {booktitle} {Advances in neural information
  processing systems}}}\ (\bibinfo {year} {2014})\ pp.\ \bibinfo {pages}
  {2672--2680}\BibitemShut {NoStop}%
\bibitem [{\citenamefont {Bengio}\ \emph {et~al.}(2009)\citenamefont {Bengio}
  \emph {et~al.}}]{bengio2009learning}%
  \BibitemOpen
  \bibfield  {author} {\bibinfo {author} {\bibfnamefont {Yoshua}\ \bibnamefont
  {Bengio}} \emph {et~al.},\ }\bibfield  {title} {\enquote {\bibinfo {title}
  {Learning deep architectures for ai},}\ }\href@noop {} {\bibfield  {journal}
  {\bibinfo  {journal} {Foundations and trend in Machine Learning}\ }\textbf
  {\bibinfo {volume} {2}},\ \bibinfo {pages} {1--127} (\bibinfo {year}
  {2009})}\BibitemShut {NoStop}%
\bibitem [{\citenamefont {Roth}(1996)}]{roth1996hardness}%
  \BibitemOpen
  \bibfield  {author} {\bibinfo {author} {\bibfnamefont {Dan}\ \bibnamefont
  {Roth}},\ }\bibfield  {title} {\enquote {\bibinfo {title} {On the hardness of
  approximate reasoning},}\ }\href@noop {} {\bibfield  {journal} {\bibinfo
  {journal} {Artificial Intelligence}\ }\textbf {\bibinfo {volume} {82}},\
  \bibinfo {pages} {273--302} (\bibinfo {year} {1996})}\BibitemShut {NoStop}%
\bibitem [{\citenamefont {Bengio}\ \emph
  {et~al.}(2013{\natexlab{b}})\citenamefont {Bengio}, \citenamefont {Mesnil},
  \citenamefont {Dauphin},\ and\ \citenamefont {Rifai}}]{bengio2013better}%
  \BibitemOpen
  \bibfield  {author} {\bibinfo {author} {\bibfnamefont {Yoshua}\ \bibnamefont
  {Bengio}}, \bibinfo {author} {\bibfnamefont {Gr{\'e}goire}\ \bibnamefont
  {Mesnil}}, \bibinfo {author} {\bibfnamefont {Yann}\ \bibnamefont {Dauphin}},
  \ and\ \bibinfo {author} {\bibfnamefont {Salah}\ \bibnamefont {Rifai}},\
  }\bibfield  {title} {\enquote {\bibinfo {title} {Better mixing via deep
  representations},}\ }in\ \href@noop {} {\emph {\bibinfo {booktitle}
  {Proceedings of the 30th International Conference on Machine Learning
  (ICML-13)}}}\ (\bibinfo {year} {2013})\ pp.\ \bibinfo {pages}
  {552--560}\BibitemShut {NoStop}%
\bibitem [{\citenamefont {Dumoulin}\ \emph {et~al.}(2014)\citenamefont
  {Dumoulin}, \citenamefont {Goodfellow}, \citenamefont {Courville},\ and\
  \citenamefont {Bengio}}]{Dumolin-2014}%
  \BibitemOpen
  \bibfield  {author} {\bibinfo {author} {\bibfnamefont {V.}~\bibnamefont
  {Dumoulin}}, \bibinfo {author} {\bibfnamefont {I.~J.}\ \bibnamefont
  {Goodfellow}}, \bibinfo {author} {\bibfnamefont {A.~C.}\ \bibnamefont
  {Courville}}, \ and\ \bibinfo {author} {\bibfnamefont {Y.}~\bibnamefont
  {Bengio}},\ }\bibfield  {title} {\enquote {\bibinfo {title} {On the
  challenges of physical implementations of {RBM}s},}\ }in\ \href@noop {}
  {\emph {\bibinfo {booktitle} {Proceedings of the Twenty-Eighth {AAAI}
  Conference on Artificial Intelligence, July 27 -31, 2014, Qu{\'{e}}bec City,
  Qu{\'{e}}bec, Canada.}}}\ (\bibinfo {year} {2014})\ pp.\ \bibinfo {pages}
  {1199--1205}\BibitemShut {NoStop}%
\bibitem [{\citenamefont {Korenkevych}\ \emph {et~al.}(2016)\citenamefont
  {Korenkevych}, \citenamefont {Xue}, \citenamefont {Bian}, \citenamefont
  {Chudak}, \citenamefont {Macready}, \citenamefont {Rolfe},\ and\
  \citenamefont {Andriyash}}]{korenkevych2016benchmarking}%
  \BibitemOpen
  \bibfield  {author} {\bibinfo {author} {\bibfnamefont {Dmytro}\ \bibnamefont
  {Korenkevych}}, \bibinfo {author} {\bibfnamefont {Yanbo}\ \bibnamefont
  {Xue}}, \bibinfo {author} {\bibfnamefont {Zhengbing}\ \bibnamefont {Bian}},
  \bibinfo {author} {\bibfnamefont {Fabian}\ \bibnamefont {Chudak}}, \bibinfo
  {author} {\bibfnamefont {William~G}\ \bibnamefont {Macready}}, \bibinfo
  {author} {\bibfnamefont {Jason}\ \bibnamefont {Rolfe}}, \ and\ \bibinfo
  {author} {\bibfnamefont {Evgeny}\ \bibnamefont {Andriyash}},\ }\bibfield
  {title} {\enquote {\bibinfo {title} {Benchmarking quantum hardware for
  training of fully visible boltzmann machines},}\ }\href@noop {} {\bibfield
  {journal} {\bibinfo  {journal} {arXiv preprint arXiv:1611.04528}\ } (\bibinfo
  {year} {2016})}\BibitemShut {NoStop}%
\bibitem [{\citenamefont {Chowdhury}\ and\ \citenamefont
  {Somma}(2016)}]{chowdhury2016quantum}%
  \BibitemOpen
  \bibfield  {author} {\bibinfo {author} {\bibfnamefont {Anirban~Narayan}\
  \bibnamefont {Chowdhury}}\ and\ \bibinfo {author} {\bibfnamefont {Rolando~D}\
  \bibnamefont {Somma}},\ }\bibfield  {title} {\enquote {\bibinfo {title}
  {Quantum algorithms for gibbs sampling and hitting-time estimation},}\
  }\href@noop {} {\bibfield  {journal} {\bibinfo  {journal} {arXiv preprint
  arXiv:1603.02940}\ } (\bibinfo {year} {2016})}\BibitemShut {NoStop}%
\bibitem [{\citenamefont {LeCun}\ \emph {et~al.}(2015)\citenamefont {LeCun},
  \citenamefont {Bengio},\ and\ \citenamefont {Hinton}}]{LeCun-Nature-2015}%
  \BibitemOpen
  \bibfield  {author} {\bibinfo {author} {\bibfnamefont {Yann}\ \bibnamefont
  {LeCun}}, \bibinfo {author} {\bibfnamefont {Yoshua}\ \bibnamefont {Bengio}},
  \ and\ \bibinfo {author} {\bibfnamefont {Geoffrey}\ \bibnamefont {Hinton}},\
  }\bibfield  {title} {\enquote {\bibinfo {title} {{Deep learning}},}\
  }\href@noop {} {\bibfield  {journal} {\bibinfo  {journal} {Nature}\ }\textbf
  {\bibinfo {volume} {521}},\ \bibinfo {pages} {436 -- 444} (\bibinfo {year}
  {2015})}\BibitemShut {NoStop}%
\bibitem [{\citenamefont {Lake}\ \emph {et~al.}(2015)\citenamefont {Lake},
  \citenamefont {Salakhutdinov},\ and\ \citenamefont
  {Tenenbaum}}]{lake2015human}%
  \BibitemOpen
  \bibfield  {author} {\bibinfo {author} {\bibfnamefont {Brenden~M}\
  \bibnamefont {Lake}}, \bibinfo {author} {\bibfnamefont {Ruslan}\ \bibnamefont
  {Salakhutdinov}}, \ and\ \bibinfo {author} {\bibfnamefont {Joshua~B}\
  \bibnamefont {Tenenbaum}},\ }\bibfield  {title} {\enquote {\bibinfo {title}
  {Human-level concept learning through probabilistic program induction},}\
  }\href@noop {} {\bibfield  {journal} {\bibinfo  {journal} {Science}\ }\textbf
  {\bibinfo {volume} {350}},\ \bibinfo {pages} {1332--1338} (\bibinfo {year}
  {2015})}\BibitemShut {NoStop}%
\bibitem [{\citenamefont {Boixo}\ \emph {et~al.}(2016)\citenamefont {Boixo},
  \citenamefont {Isakov}, \citenamefont {Smelyanskiy}, \citenamefont {Babbush},
  \citenamefont {Ding}, \citenamefont {Jiang}, \citenamefont {Martinis},\ and\
  \citenamefont {Neven}}]{boixo2016characterizing}%
  \BibitemOpen
  \bibfield  {author} {\bibinfo {author} {\bibfnamefont {Sergio}\ \bibnamefont
  {Boixo}}, \bibinfo {author} {\bibfnamefont {Sergei~V}\ \bibnamefont
  {Isakov}}, \bibinfo {author} {\bibfnamefont {Vadim~N}\ \bibnamefont
  {Smelyanskiy}}, \bibinfo {author} {\bibfnamefont {Ryan}\ \bibnamefont
  {Babbush}}, \bibinfo {author} {\bibfnamefont {Nan}\ \bibnamefont {Ding}},
  \bibinfo {author} {\bibfnamefont {Zhang}\ \bibnamefont {Jiang}}, \bibinfo
  {author} {\bibfnamefont {John~M}\ \bibnamefont {Martinis}}, \ and\ \bibinfo
  {author} {\bibfnamefont {Hartmut}\ \bibnamefont {Neven}},\ }\bibfield
  {title} {\enquote {\bibinfo {title} {Characterizing quantum supremacy in
  near-term devices},}\ }\href@noop {} {\bibfield  {journal} {\bibinfo
  {journal} {arXiv preprint arXiv:1608.00263}\ } (\bibinfo {year}
  {2016})}\BibitemShut {NoStop}%
\bibitem [{\citenamefont {Gu}\ \emph {et~al.}(2012)\citenamefont {Gu},
  \citenamefont {Wiesner}, \citenamefont {Rieper},\ and\ \citenamefont
  {Vedral}}]{Gu2012}%
  \BibitemOpen
  \bibfield  {author} {\bibinfo {author} {\bibfnamefont {Mile}\ \bibnamefont
  {Gu}}, \bibinfo {author} {\bibfnamefont {Karoline}\ \bibnamefont {Wiesner}},
  \bibinfo {author} {\bibfnamefont {Elisabeth}\ \bibnamefont {Rieper}}, \ and\
  \bibinfo {author} {\bibfnamefont {Vlatko}\ \bibnamefont {Vedral}},\
  }\bibfield  {title} {\enquote {\bibinfo {title} {Quantum mechanics can reduce
  the complexity of classical models},}\ }\href@noop {} {\bibfield  {journal}
  {\bibinfo  {journal} {Nature Communications}\ }\textbf {\bibinfo {volume}
  {3}},\ \bibinfo {pages} {762} (\bibinfo {year} {2012})}\BibitemShut {NoStop}%
\bibitem [{\citenamefont {Palsson}\ \emph {et~al.}(2017)\citenamefont
  {Palsson}, \citenamefont {Gu}, \citenamefont {Ho}, \citenamefont {Wiseman},\
  and\ \citenamefont {Pryde}}]{palsson2017experimentally}%
  \BibitemOpen
  \bibfield  {author} {\bibinfo {author} {\bibfnamefont {Matthew~S}\
  \bibnamefont {Palsson}}, \bibinfo {author} {\bibfnamefont {Mile}\
  \bibnamefont {Gu}}, \bibinfo {author} {\bibfnamefont {Joseph}\ \bibnamefont
  {Ho}}, \bibinfo {author} {\bibfnamefont {Howard~M}\ \bibnamefont {Wiseman}},
  \ and\ \bibinfo {author} {\bibfnamefont {Geoff~J}\ \bibnamefont {Pryde}},\
  }\bibfield  {title} {\enquote {\bibinfo {title} {Experimentally modeling
  stochastic processes with less memory by the use of a quantum processor},}\
  }\href@noop {} {\bibfield  {journal} {\bibinfo  {journal} {Science Advances}\
  }\textbf {\bibinfo {volume} {3}},\ \bibinfo {pages} {e1601302} (\bibinfo
  {year} {2017})}\BibitemShut {NoStop}%
\bibitem [{\citenamefont {Elliott}\ and\ \citenamefont
  {Gu}(2017)}]{elliott2017occam}%
  \BibitemOpen
  \bibfield  {author} {\bibinfo {author} {\bibfnamefont {Thomas~J}\
  \bibnamefont {Elliott}}\ and\ \bibinfo {author} {\bibfnamefont {Mile}\
  \bibnamefont {Gu}},\ }\bibfield  {title} {\enquote {\bibinfo {title} {Occam's
  vorpal quantum razor: Memory reduction when simulating continuous-time
  stochastic processes with quantum devices},}\ }\href@noop {} {\bibfield
  {journal} {\bibinfo  {journal} {arXiv preprint arXiv:1704.04231}\ } (\bibinfo
  {year} {2017})}\BibitemShut {NoStop}%
\bibitem [{\citenamefont {Burnham}\ and\ \citenamefont
  {Anderson}(2003)}]{burnham2003model}%
  \BibitemOpen
  \bibfield  {author} {\bibinfo {author} {\bibfnamefont {Kenneth~P}\
  \bibnamefont {Burnham}}\ and\ \bibinfo {author} {\bibfnamefont {David~R}\
  \bibnamefont {Anderson}},\ }\href@noop {} {\emph {\bibinfo {title} {Model
  selection and multimodel inference: a practical information-theoretic
  approach}}}\ (\bibinfo  {publisher} {Springer Science \& Business Media},\
  \bibinfo {year} {2003})\BibitemShut {NoStop}%
\bibitem [{\citenamefont {Jaynes}(1957)}]{Jaynes-PhysRev-1957}%
  \BibitemOpen
  \bibfield  {author} {\bibinfo {author} {\bibfnamefont {E.~T.}\ \bibnamefont
  {Jaynes}},\ }\bibfield  {title} {\enquote {\bibinfo {title} {Information
  theory and statistical mechanics},}\ }\href {\doibase
  10.1103/PhysRev.106.620} {\bibfield  {journal} {\bibinfo  {journal} {Phys.
  Rev.}\ }\textbf {\bibinfo {volume} {106}},\ \bibinfo {pages} {620--630}
  (\bibinfo {year} {1957})}\BibitemShut {NoStop}%
\bibitem [{\citenamefont {Perc}\ \emph {et~al.}(2017)\citenamefont {Perc},
  \citenamefont {Jordan}, \citenamefont {Rand}, \citenamefont {Wang},
  \citenamefont {Boccaletti},\ and\ \citenamefont
  {Szolnoki}}]{perc2017statistical}%
  \BibitemOpen
  \bibfield  {author} {\bibinfo {author} {\bibfnamefont {Matja{\v{z}}}\
  \bibnamefont {Perc}}, \bibinfo {author} {\bibfnamefont {Jillian~J}\
  \bibnamefont {Jordan}}, \bibinfo {author} {\bibfnamefont {David~G}\
  \bibnamefont {Rand}}, \bibinfo {author} {\bibfnamefont {Zhen}\ \bibnamefont
  {Wang}}, \bibinfo {author} {\bibfnamefont {Stefano}\ \bibnamefont
  {Boccaletti}}, \ and\ \bibinfo {author} {\bibfnamefont {Attila}\ \bibnamefont
  {Szolnoki}},\ }\bibfield  {title} {\enquote {\bibinfo {title} {Statistical
  physics of human cooperation},}\ }\href@noop {} {\bibfield  {journal}
  {\bibinfo  {journal} {Physics Reports}\ } (\bibinfo {year}
  {2017})}\BibitemShut {NoStop}%
\bibitem [{\citenamefont {Realpe-G{\'o}mez}\ \emph {et~al.}(2018)\citenamefont
  {Realpe-G{\'o}mez}, \citenamefont {Andrighetto}, \citenamefont {Nardin},\
  and\ \citenamefont {Montoya}}]{realpe2016balancing}%
  \BibitemOpen
  \bibfield  {author} {\bibinfo {author} {\bibfnamefont {John}\ \bibnamefont
  {Realpe-G{\'o}mez}}, \bibinfo {author} {\bibfnamefont {Giulia}\ \bibnamefont
  {Andrighetto}}, \bibinfo {author} {\bibfnamefont {Gustavo}\ \bibnamefont
  {Nardin}}, \ and\ \bibinfo {author} {\bibfnamefont {Javier~Antonio}\
  \bibnamefont {Montoya}},\ }\bibfield  {title} {\enquote {\bibinfo {title}
  {Balancing selfishness and norm conformity can explain human behavior in
  large-scale prisoners dilemma games and can poise human groups near
  criticality},}\ }\href@noop {} {\bibfield  {journal} {\bibinfo  {journal}
  {Physical Review E, to appear}\ } (\bibinfo {year} {2018})},\ \bibinfo {note}
  {arXiv preprint arXiv:1608.01291}\BibitemShut {NoStop}%
\bibitem [{\citenamefont {M{\'e}zard}\ \emph {et~al.}(2002)\citenamefont
  {M{\'e}zard}, \citenamefont {Parisi},\ and\ \citenamefont
  {Zecchina}}]{mezard2002analytic}%
  \BibitemOpen
  \bibfield  {author} {\bibinfo {author} {\bibfnamefont {Marc}\ \bibnamefont
  {M{\'e}zard}}, \bibinfo {author} {\bibfnamefont {Giorgio}\ \bibnamefont
  {Parisi}}, \ and\ \bibinfo {author} {\bibfnamefont {Riccardo}\ \bibnamefont
  {Zecchina}},\ }\bibfield  {title} {\enquote {\bibinfo {title} {Analytic and
  algorithmic solution of random satisfiability problems},}\ }\href@noop {}
  {\bibfield  {journal} {\bibinfo  {journal} {Science}\ }\textbf {\bibinfo
  {volume} {297}},\ \bibinfo {pages} {812--815} (\bibinfo {year}
  {2002})}\BibitemShut {NoStop}%
\bibitem [{\citenamefont {Mezard}\ and\ \citenamefont
  {Montanari}(2009)}]{Mezard-book-2009}%
  \BibitemOpen
  \bibfield  {author} {\bibinfo {author} {\bibfnamefont {Marc}\ \bibnamefont
  {Mezard}}\ and\ \bibinfo {author} {\bibfnamefont {Andrea}\ \bibnamefont
  {Montanari}},\ }\href@noop {} {\emph {\bibinfo {title} {Information, Physics,
  and Computation}}}\ (\bibinfo  {publisher} {Oxford University Press, Inc.},\
  \bibinfo {address} {New York, NY, USA},\ \bibinfo {year} {2009})\BibitemShut
  {NoStop}%
\bibitem [{\citenamefont {D'Ariano}\ \emph {et~al.}(2017)\citenamefont
  {D'Ariano}, \citenamefont {Chiribella},\ and\ \citenamefont
  {Perinotti}}]{d2017quantum}%
  \BibitemOpen
  \bibfield  {author} {\bibinfo {author} {\bibfnamefont {Giacomo~Mauro}\
  \bibnamefont {D'Ariano}}, \bibinfo {author} {\bibfnamefont {Giulio}\
  \bibnamefont {Chiribella}}, \ and\ \bibinfo {author} {\bibfnamefont {Paolo}\
  \bibnamefont {Perinotti}},\ }\href@noop {} {\emph {\bibinfo {title} {Quantum
  Theory from First Principles: An Informational Approach}}}\ (\bibinfo
  {publisher} {Cambridge University Press},\ \bibinfo {year}
  {2017})\BibitemShut {NoStop}%
\bibitem [{\citenamefont {Realpe-G{\'o}mez}(2017)}]{realpe2017quantum}%
  \BibitemOpen
  \bibfield  {author} {\bibinfo {author} {\bibfnamefont {John}\ \bibnamefont
  {Realpe-G{\'o}mez}},\ }\bibfield  {title} {\enquote {\bibinfo {title}
  {Quantum as self-reference},}\ }\href@noop {} {\bibfield  {journal} {\bibinfo
   {journal} {arXiv preprint arXiv:1705.04307}\ } (\bibinfo {year}
  {2017})}\BibitemShut {NoStop}%
\bibitem [{\citenamefont {Bruza}\ \emph {et~al.}(2015)\citenamefont {Bruza},
  \citenamefont {Wang},\ and\ \citenamefont {Busemeyer}}]{bruza2015quantum}%
  \BibitemOpen
  \bibfield  {author} {\bibinfo {author} {\bibfnamefont {Peter~D}\ \bibnamefont
  {Bruza}}, \bibinfo {author} {\bibfnamefont {Zheng}\ \bibnamefont {Wang}}, \
  and\ \bibinfo {author} {\bibfnamefont {Jerome~R}\ \bibnamefont {Busemeyer}},\
  }\bibfield  {title} {\enquote {\bibinfo {title} {Quantum cognition: a new
  theoretical approach to psychology},}\ }\href@noop {} {\bibfield  {journal}
  {\bibinfo  {journal} {Trends in cognitive sciences}\ }\textbf {\bibinfo
  {volume} {19}},\ \bibinfo {pages} {383--393} (\bibinfo {year}
  {2015})}\BibitemShut {NoStop}%
\bibitem [{\citenamefont {Busemeyer}\ and\ \citenamefont
  {Bruza}(2012)}]{busemeyer2012quantum}%
  \BibitemOpen
  \bibfield  {author} {\bibinfo {author} {\bibfnamefont {Jerome~R}\
  \bibnamefont {Busemeyer}}\ and\ \bibinfo {author} {\bibfnamefont {Peter~D}\
  \bibnamefont {Bruza}},\ }\href@noop {} {\emph {\bibinfo {title} {Quantum
  models of cognition and decision}}}\ (\bibinfo  {publisher} {Cambridge
  University Press},\ \bibinfo {year} {2012})\BibitemShut {NoStop}%
\bibitem [{\citenamefont {Aerts}\ \emph {et~al.}(2016)\citenamefont {Aerts},
  \citenamefont {Broekaert}, \citenamefont {Gabora},\ and\ \citenamefont
  {Sozzo}}]{aerts2016quantum}%
  \BibitemOpen
  \bibfield  {author} {\bibinfo {author} {\bibfnamefont {Diederik}\
  \bibnamefont {Aerts}}, \bibinfo {author} {\bibfnamefont {Jan}\ \bibnamefont
  {Broekaert}}, \bibinfo {author} {\bibfnamefont {Liane}\ \bibnamefont
  {Gabora}}, \ and\ \bibinfo {author} {\bibfnamefont {Sandro}\ \bibnamefont
  {Sozzo}},\ }\bibfield  {title} {\enquote {\bibinfo {title} {Quantum
  structures in cognitive and social science},}\ }\href@noop {} {\bibfield
  {journal} {\bibinfo  {journal} {Frontiers in psychology}\ }\textbf {\bibinfo
  {volume} {7}} (\bibinfo {year} {2016})}\BibitemShut {NoStop}%
\bibitem [{\citenamefont {Busemeyer}\ \emph {et~al.}(2015)\citenamefont
  {Busemeyer}, \citenamefont {Wang},\ and\ \citenamefont
  {Shiffrin}}]{busemeyer2015bayesian}%
  \BibitemOpen
  \bibfield  {author} {\bibinfo {author} {\bibfnamefont {Jerome~R}\
  \bibnamefont {Busemeyer}}, \bibinfo {author} {\bibfnamefont {Zheng}\
  \bibnamefont {Wang}}, \ and\ \bibinfo {author} {\bibfnamefont {Richard~M}\
  \bibnamefont {Shiffrin}},\ }\bibfield  {title} {\enquote {\bibinfo {title}
  {Bayesian model comparison favors quantum over standard decision theory
  account of dynamic inconsistency.}}\ }\href@noop {} {\bibfield  {journal}
  {\bibinfo  {journal} {Decision}\ }\textbf {\bibinfo {volume} {2}},\ \bibinfo
  {pages} {1} (\bibinfo {year} {2015})}\BibitemShut {NoStop}%
\bibitem [{\citenamefont {Wang}\ \emph {et~al.}(2014)\citenamefont {Wang},
  \citenamefont {Solloway}, \citenamefont {Shiffrin},\ and\ \citenamefont
  {Busemeyer}}]{wang2014context}%
  \BibitemOpen
  \bibfield  {author} {\bibinfo {author} {\bibfnamefont {Zheng}\ \bibnamefont
  {Wang}}, \bibinfo {author} {\bibfnamefont {Tyler}\ \bibnamefont {Solloway}},
  \bibinfo {author} {\bibfnamefont {Richard~M}\ \bibnamefont {Shiffrin}}, \
  and\ \bibinfo {author} {\bibfnamefont {Jerome~R}\ \bibnamefont {Busemeyer}},\
  }\bibfield  {title} {\enquote {\bibinfo {title} {Context effects produced by
  question orders reveal quantum nature of human judgments},}\ }\href@noop {}
  {\bibfield  {journal} {\bibinfo  {journal} {Proceedings of the National
  Academy of Sciences}\ }\textbf {\bibinfo {volume} {111}},\ \bibinfo {pages}
  {9431--9436} (\bibinfo {year} {2014})}\BibitemShut {NoStop}%
\bibitem [{\citenamefont {Gracia-L{\'a}zaro}\ \emph {et~al.}(2012)\citenamefont
  {Gracia-L{\'a}zaro}, \citenamefont {Ferrer}, \citenamefont {Ruiz},
  \citenamefont {Taranc{\'o}n}, \citenamefont {Cuesta}, \citenamefont
  {S{\'a}nchez},\ and\ \citenamefont {Moreno}}]{gracia2012heterogeneous}%
  \BibitemOpen
  \bibfield  {author} {\bibinfo {author} {\bibfnamefont {Carlos}\ \bibnamefont
  {Gracia-L{\'a}zaro}}, \bibinfo {author} {\bibfnamefont {Alfredo}\
  \bibnamefont {Ferrer}}, \bibinfo {author} {\bibfnamefont {Gonzalo}\
  \bibnamefont {Ruiz}}, \bibinfo {author} {\bibfnamefont {Alfonso}\
  \bibnamefont {Taranc{\'o}n}}, \bibinfo {author} {\bibfnamefont {Jos{\'e}~A}\
  \bibnamefont {Cuesta}}, \bibinfo {author} {\bibfnamefont {Angel}\
  \bibnamefont {S{\'a}nchez}}, \ and\ \bibinfo {author} {\bibfnamefont {Yamir}\
  \bibnamefont {Moreno}},\ }\bibfield  {title} {\enquote {\bibinfo {title}
  {Heterogeneous networks do not promote cooperation when humans play a
  prisonerâ€™s dilemma},}\ }\href@noop {} {\bibfield  {journal} {\bibinfo
  {journal} {Proceedings of the National Academy of Sciences}\ }\textbf
  {\bibinfo {volume} {109}},\ \bibinfo {pages} {12922--12926} (\bibinfo {year}
  {2012})}\BibitemShut {NoStop}%
\bibitem [{\citenamefont {Guti{\'e}rrez-Roig}\ \emph
  {et~al.}(2014)\citenamefont {Guti{\'e}rrez-Roig}, \citenamefont
  {Gracia-L{\'a}zaro}, \citenamefont {Perell{\'o}}, \citenamefont {Moreno},\
  and\ \citenamefont {S{\'a}nchez}}]{gutierrez2014transition}%
  \BibitemOpen
  \bibfield  {author} {\bibinfo {author} {\bibfnamefont {Mario}\ \bibnamefont
  {Guti{\'e}rrez-Roig}}, \bibinfo {author} {\bibfnamefont {Carlos}\
  \bibnamefont {Gracia-L{\'a}zaro}}, \bibinfo {author} {\bibfnamefont {Josep}\
  \bibnamefont {Perell{\'o}}}, \bibinfo {author} {\bibfnamefont {Yamir}\
  \bibnamefont {Moreno}}, \ and\ \bibinfo {author} {\bibfnamefont {Angel}\
  \bibnamefont {S{\'a}nchez}},\ }\bibfield  {title} {\enquote {\bibinfo {title}
  {Transition from reciprocal cooperation to persistent behaviour in social
  dilemmas at the end of adolescence},}\ }\href@noop {} {\bibfield  {journal}
  {\bibinfo  {journal} {Nature communications}\ }\textbf {\bibinfo {volume}
  {5}} (\bibinfo {year} {2014})}\BibitemShut {NoStop}%
\bibitem [{\citenamefont {House}(2016)}]{house2016artificial}%
  \BibitemOpen
  \bibfield  {author} {\bibinfo {author} {\bibfnamefont {White}\ \bibnamefont
  {House}},\ }\bibfield  {title} {\enquote {\bibinfo {title} {Artificial
  intelligence, automation, and the economy},}\ }\href@noop {} {\bibfield
  {journal} {\bibinfo  {journal} {Executive office of the President.
  https://obamawhitehouse. archives. gov/sites/whitehouse.
  gov/files/documents/Artificial-Intelligence-Automation-Economy. PDF}\ }
  (\bibinfo {year} {2016})}\BibitemShut {NoStop}%
\bibitem [{\citenamefont {Harrow}\ \emph {et~al.}(2009)\citenamefont {Harrow},
  \citenamefont {Hassidim},\ and\ \citenamefont {Lloyd}}]{harrow2009quantum}%
  \BibitemOpen
  \bibfield  {author} {\bibinfo {author} {\bibfnamefont {Aram~W}\ \bibnamefont
  {Harrow}}, \bibinfo {author} {\bibfnamefont {Avinatan}\ \bibnamefont
  {Hassidim}}, \ and\ \bibinfo {author} {\bibfnamefont {Seth}\ \bibnamefont
  {Lloyd}},\ }\bibfield  {title} {\enquote {\bibinfo {title} {Quantum algorithm
  for linear systems of equations},}\ }\href@noop {} {\bibfield  {journal}
  {\bibinfo  {journal} {Physical review letters}\ }\textbf {\bibinfo {volume}
  {103}},\ \bibinfo {pages} {150502} (\bibinfo {year} {2009})}\BibitemShut
  {NoStop}%
\bibitem [{\citenamefont {Hinton}(2002)}]{hinton2002training}%
  \BibitemOpen
  \bibfield  {author} {\bibinfo {author} {\bibfnamefont {Geoffrey~E}\
  \bibnamefont {Hinton}},\ }\bibfield  {title} {\enquote {\bibinfo {title}
  {Training products of experts by minimizing contrastive divergence},}\
  }\href@noop {} {\bibfield  {journal} {\bibinfo  {journal} {{Neural
  Computation}}\ }\textbf {\bibinfo {volume} {14}},\ \bibinfo {pages}
  {1771--1800} (\bibinfo {year} {2002})}\BibitemShut {NoStop}%
\bibitem [{\citenamefont {Carreira-Perpinan}\ and\ \citenamefont
  {Hinton}(2005)}]{carreira2005contrastive}%
  \BibitemOpen
  \bibfield  {author} {\bibinfo {author} {\bibfnamefont {Miguel~A}\
  \bibnamefont {Carreira-Perpinan}}\ and\ \bibinfo {author} {\bibfnamefont
  {Geoffrey~E}\ \bibnamefont {Hinton}},\ }\bibfield  {title} {\enquote
  {\bibinfo {title} {On contrastive divergence learning.}}\ }in\ \href@noop {}
  {\emph {\bibinfo {booktitle} {Aistats}}},\ Vol.~\bibinfo {volume} {10}\
  (\bibinfo {year} {2005})\ pp.\ \bibinfo {pages} {33--40}\BibitemShut
  {NoStop}%
\bibitem [{\citenamefont {Hyv{\"a}rinen}(2006)}]{hyvarinen2006consistency}%
  \BibitemOpen
  \bibfield  {author} {\bibinfo {author} {\bibfnamefont {Aapo}\ \bibnamefont
  {Hyv{\"a}rinen}},\ }\bibfield  {title} {\enquote {\bibinfo {title}
  {Consistency of pseudolikelihood estimation of fully visible boltzmann
  machines},}\ }\href@noop {} {\bibfield  {journal} {\bibinfo  {journal}
  {Neural Computation}\ }\textbf {\bibinfo {volume} {18}},\ \bibinfo {pages}
  {2283--2292} (\bibinfo {year} {2006})}\BibitemShut {NoStop}%
\bibitem [{\citenamefont {Raymond}\ \emph {et~al.}(2016)\citenamefont
  {Raymond}, \citenamefont {Yarkoni},\ and\ \citenamefont
  {Andriyash}}]{Raymond-DWave-2016}%
  \BibitemOpen
  \bibfield  {author} {\bibinfo {author} {\bibfnamefont {Jack}\ \bibnamefont
  {Raymond}}, \bibinfo {author} {\bibfnamefont {Sheir}\ \bibnamefont
  {Yarkoni}}, \ and\ \bibinfo {author} {\bibfnamefont {Evgeny}\ \bibnamefont
  {Andriyash}},\ }\bibfield  {title} {\enquote {\bibinfo {title} {{Global
  warming: Temperature estimation in annealers}},}\ }\href@noop {} {\bibfield
  {journal} {\bibinfo  {journal} {arXiv:1606.00919}\ } (\bibinfo {year}
  {2016})}\BibitemShut {NoStop}%
\bibitem [{\citenamefont {Beals}\ \emph {et~al.}(2013)\citenamefont {Beals},
  \citenamefont {Brierley}, \citenamefont {Gray}, \citenamefont {Harrow},
  \citenamefont {Kutin}, \citenamefont {Linden}, \citenamefont {Shepherd},\
  and\ \citenamefont {Stather}}]{beals2013efficient}%
  \BibitemOpen
  \bibfield  {author} {\bibinfo {author} {\bibfnamefont {Robert}\ \bibnamefont
  {Beals}}, \bibinfo {author} {\bibfnamefont {Stephen}\ \bibnamefont
  {Brierley}}, \bibinfo {author} {\bibfnamefont {Oliver}\ \bibnamefont {Gray}},
  \bibinfo {author} {\bibfnamefont {Aram~W}\ \bibnamefont {Harrow}}, \bibinfo
  {author} {\bibfnamefont {Samuel}\ \bibnamefont {Kutin}}, \bibinfo {author}
  {\bibfnamefont {Noah}\ \bibnamefont {Linden}}, \bibinfo {author}
  {\bibfnamefont {Dan}\ \bibnamefont {Shepherd}}, \ and\ \bibinfo {author}
  {\bibfnamefont {Mark}\ \bibnamefont {Stather}},\ }\bibfield  {title}
  {\enquote {\bibinfo {title} {Efficient distributed quantum computing},}\ }in\
  \href@noop {} {\emph {\bibinfo {booktitle} {Proc. R. Soc. A}}},\ Vol.\
  \bibinfo {volume} {469}\ (\bibinfo {organization} {The Royal Society},\
  \bibinfo {year} {2013})\ p.\ \bibinfo {pages} {20120686}\BibitemShut
  {NoStop}%
\bibitem [{\citenamefont {Choi}(2011)}]{choi2011minor}%
  \BibitemOpen
  \bibfield  {author} {\bibinfo {author} {\bibfnamefont {Vicky}\ \bibnamefont
  {Choi}},\ }\bibfield  {title} {\enquote {\bibinfo {title} {Minor-embedding in
  adiabatic quantum computation: Ii. minor-universal graph design},}\
  }\href@noop {} {\bibfield  {journal} {\bibinfo  {journal} {Quantum
  Information Processing}\ }\textbf {\bibinfo {volume} {10}},\ \bibinfo {pages}
  {343--353} (\bibinfo {year} {2011})}\BibitemShut {NoStop}%
\bibitem [{\citenamefont {Perdomo-Ortiz}\ \emph {et~al.}(2015)\citenamefont
  {Perdomo-Ortiz}, \citenamefont {Fluegemann}, \citenamefont {Biswas},\ and\
  \citenamefont {Smelyanskiy}}]{perdomo2015performance}%
  \BibitemOpen
  \bibfield  {author} {\bibinfo {author} {\bibfnamefont {Alejandro}\
  \bibnamefont {Perdomo-Ortiz}}, \bibinfo {author} {\bibfnamefont {Joseph}\
  \bibnamefont {Fluegemann}}, \bibinfo {author} {\bibfnamefont {Rupak}\
  \bibnamefont {Biswas}}, \ and\ \bibinfo {author} {\bibfnamefont {Vadim~N}\
  \bibnamefont {Smelyanskiy}},\ }\bibfield  {title} {\enquote {\bibinfo {title}
  {{A performance estimator for quantum annealers: gauge selection and
  parameter setting}},}\ }\href@noop {} {\bibfield  {journal} {\bibinfo
  {journal} {arXiv:1503.01083}\ } (\bibinfo {year} {2015})}\BibitemShut
  {NoStop}%
\bibitem [{\citenamefont {Pudenz}(2016)}]{Pudenz2016}%
  \BibitemOpen
  \bibfield  {author} {\bibinfo {author} {\bibfnamefont {Kristen~L.}\
  \bibnamefont {Pudenz}},\ }\bibfield  {title} {\enquote {\bibinfo {title}
  {Parameter setting for quantum annealers},}\ }\href@noop {} {\bibfield
  {journal} {\bibinfo  {journal} {arXiv:1611.07552}\ } (\bibinfo {year}
  {2016})}\BibitemShut {NoStop}%
\bibitem [{\citenamefont {Linke}\ \emph {et~al.}(2017)\citenamefont {Linke},
  \citenamefont {Maslov}, \citenamefont {Roetteler}, \citenamefont {Debnath},
  \citenamefont {Figgatt}, \citenamefont {Landsman}, \citenamefont {Wright},\
  and\ \citenamefont {Monroe}}]{linke2017experimental}%
  \BibitemOpen
  \bibfield  {author} {\bibinfo {author} {\bibfnamefont {Norbert~M}\
  \bibnamefont {Linke}}, \bibinfo {author} {\bibfnamefont {Dmitri}\
  \bibnamefont {Maslov}}, \bibinfo {author} {\bibfnamefont {Martin}\
  \bibnamefont {Roetteler}}, \bibinfo {author} {\bibfnamefont {Shantanu}\
  \bibnamefont {Debnath}}, \bibinfo {author} {\bibfnamefont {Caroline}\
  \bibnamefont {Figgatt}}, \bibinfo {author} {\bibfnamefont {Kevin~A}\
  \bibnamefont {Landsman}}, \bibinfo {author} {\bibfnamefont {Kenneth}\
  \bibnamefont {Wright}}, \ and\ \bibinfo {author} {\bibfnamefont
  {Christopher}\ \bibnamefont {Monroe}},\ }\bibfield  {title} {\enquote
  {\bibinfo {title} {Experimental comparison of two quantum computing
  architectures},}\ }\href@noop {} {\bibfield  {journal} {\bibinfo  {journal}
  {Proceedings of the National Academy of Sciences}\ ,\ \bibinfo {pages}
  {201618020}} (\bibinfo {year} {2017})}\BibitemShut {NoStop}%
\bibitem [{\citenamefont {Lloyd}\ and\ \citenamefont
  {Braunstein}(1999)}]{lloyd1999quantum}%
  \BibitemOpen
  \bibfield  {author} {\bibinfo {author} {\bibfnamefont {Seth}\ \bibnamefont
  {Lloyd}}\ and\ \bibinfo {author} {\bibfnamefont {Samuel~L}\ \bibnamefont
  {Braunstein}},\ }\bibfield  {title} {\enquote {\bibinfo {title} {Quantum
  computation over continuous variables},}\ }\href@noop {} {\bibfield
  {journal} {\bibinfo  {journal} {Physical Review Letters}\ }\textbf {\bibinfo
  {volume} {82}},\ \bibinfo {pages} {1784} (\bibinfo {year}
  {1999})}\BibitemShut {NoStop}%
\bibitem [{\citenamefont {Lau}\ \emph {et~al.}(2017)\citenamefont {Lau},
  \citenamefont {Pooser}, \citenamefont {Siopsis},\ and\ \citenamefont
  {Weedbrook}}]{lau2017quantum}%
  \BibitemOpen
  \bibfield  {author} {\bibinfo {author} {\bibfnamefont {Hoi-Kwan}\
  \bibnamefont {Lau}}, \bibinfo {author} {\bibfnamefont {Raphael}\ \bibnamefont
  {Pooser}}, \bibinfo {author} {\bibfnamefont {George}\ \bibnamefont
  {Siopsis}}, \ and\ \bibinfo {author} {\bibfnamefont {Christian}\ \bibnamefont
  {Weedbrook}},\ }\bibfield  {title} {\enquote {\bibinfo {title} {Quantum
  machine learning over infinite dimensions},}\ }\href@noop {} {\bibfield
  {journal} {\bibinfo  {journal} {Physical Review Letters}\ }\textbf {\bibinfo
  {volume} {118}},\ \bibinfo {pages} {080501} (\bibinfo {year}
  {2017})}\BibitemShut {NoStop}%
\bibitem [{\citenamefont {{Das}}\ \emph {et~al.}(2017)\citenamefont {{Das}},
  \citenamefont {{Siopsis}},\ and\ \citenamefont
  {{Weedbrook}}}]{2017arXiv170700360D}%
  \BibitemOpen
  \bibfield  {author} {\bibinfo {author} {\bibfnamefont {S.}~\bibnamefont
  {{Das}}}, \bibinfo {author} {\bibfnamefont {G.}~\bibnamefont {{Siopsis}}}, \
  and\ \bibinfo {author} {\bibfnamefont {C.}~\bibnamefont {{Weedbrook}}},\
  }\bibfield  {title} {\enquote {\bibinfo {title} {{Continuous-variable quantum
  Gaussian process regression and quantum singular value decomposition of
  non-sparse low rank matrices}},}\ }\href@noop {} {\bibfield  {journal}
  {\bibinfo  {journal} {ArXiv e-prints}\ } (\bibinfo {year} {2017})},\ \Eprint
  {http://arxiv.org/abs/1707.00360} {arXiv:1707.00360 [quant-ph]} \BibitemShut
  {NoStop}%
\bibitem [{\citenamefont {Hinton}\ \emph {et~al.}(1995)\citenamefont {Hinton},
  \citenamefont {Dayan}, \citenamefont {Frey},\ and\ \citenamefont
  {Neal}}]{hinton1995wake}%
  \BibitemOpen
  \bibfield  {author} {\bibinfo {author} {\bibfnamefont {Geoffrey~E.}\
  \bibnamefont {Hinton}}, \bibinfo {author} {\bibfnamefont {Peter}\
  \bibnamefont {Dayan}}, \bibinfo {author} {\bibfnamefont {Brendan~J.}\
  \bibnamefont {Frey}}, \ and\ \bibinfo {author} {\bibfnamefont {Radford~M.}\
  \bibnamefont {Neal}},\ }\bibfield  {title} {\enquote {\bibinfo {title} {The
  wake-sleep algorithm for unsupervised neural networks},}\ }\href@noop {}
  {\bibfield  {journal} {\bibinfo  {journal} {Science}\ }\textbf {\bibinfo
  {volume} {268}},\ \bibinfo {pages} {1158} (\bibinfo {year}
  {1995})}\BibitemShut {NoStop}%
\bibitem [{\citenamefont {Dayan}\ \emph {et~al.}(1995)\citenamefont {Dayan},
  \citenamefont {Hinton}, \citenamefont {Neal},\ and\ \citenamefont
  {Zemel}}]{dayan1995helmholtz}%
  \BibitemOpen
  \bibfield  {author} {\bibinfo {author} {\bibfnamefont {Peter}\ \bibnamefont
  {Dayan}}, \bibinfo {author} {\bibfnamefont {Geoffrey~E}\ \bibnamefont
  {Hinton}}, \bibinfo {author} {\bibfnamefont {Radford~M}\ \bibnamefont
  {Neal}}, \ and\ \bibinfo {author} {\bibfnamefont {Richard~S}\ \bibnamefont
  {Zemel}},\ }\bibfield  {title} {\enquote {\bibinfo {title} {The helmholtz
  machine},}\ }\href@noop {} {\bibfield  {journal} {\bibinfo  {journal} {Neural
  computation}\ }\textbf {\bibinfo {volume} {7}},\ \bibinfo {pages} {889--904}
  (\bibinfo {year} {1995})}\BibitemShut {NoStop}%
\bibitem [{\citenamefont {Dayan}\ and\ \citenamefont
  {Hinton}(1996)}]{dayan1996varieties}%
  \BibitemOpen
  \bibfield  {author} {\bibinfo {author} {\bibfnamefont {Peter}\ \bibnamefont
  {Dayan}}\ and\ \bibinfo {author} {\bibfnamefont {Geoffrey~E}\ \bibnamefont
  {Hinton}},\ }\bibfield  {title} {\enquote {\bibinfo {title} {Varieties of
  helmholtz machine},}\ }\href@noop {} {\bibfield  {journal} {\bibinfo
  {journal} {Neural Networks}\ }\textbf {\bibinfo {volume} {9}},\ \bibinfo
  {pages} {1385--1403} (\bibinfo {year} {1996})}\BibitemShut {NoStop}%
\bibitem [{\citenamefont {Kvam}\ \emph {et~al.}(2015)\citenamefont {Kvam},
  \citenamefont {Pleskac}, \citenamefont {Yu},\ and\ \citenamefont
  {Busemeyer}}]{kvam2015interference}%
  \BibitemOpen
  \bibfield  {author} {\bibinfo {author} {\bibfnamefont {Peter~D}\ \bibnamefont
  {Kvam}}, \bibinfo {author} {\bibfnamefont {Timothy~J}\ \bibnamefont
  {Pleskac}}, \bibinfo {author} {\bibfnamefont {Shuli}\ \bibnamefont {Yu}}, \
  and\ \bibinfo {author} {\bibfnamefont {Jerome~R}\ \bibnamefont {Busemeyer}},\
  }\bibfield  {title} {\enquote {\bibinfo {title} {Interference effects of
  choice on confidence: Quantum characteristics of evidence accumulation},}\
  }\href@noop {} {\bibfield  {journal} {\bibinfo  {journal} {Proceedings of the
  National Academy of Sciences}\ }\textbf {\bibinfo {volume} {112}},\ \bibinfo
  {pages} {10645--10650} (\bibinfo {year} {2015})}\BibitemShut {NoStop}%
\end{thebibliography}

%

\end{document}